\documentclass[useAMS,usenatbib,usegraphicx]{mn2e}
 
\title[The Galaxy Merger History at $z < 1.2$]{The Structures of Distant Galaxies - III: The Merger History of over $20,000$ Massive Galaxies at $z < 1.2$}
\author[Conselice, Yang and Bluck]{ Christopher J. Conselice$^{1}$\thanks{E-mail:
conselice@nottingham.ac.uk}, Cui Yang$^{1}$, Asa F. L. Bluck$^{1}$ \\
$^{1}$University of Nottingham, School of Physics \& Astronomy, Nottingham, NG7 2RD UK}
\def\deg{$^{\circ}\,$}
\def\solm{M$_{\odot}\,$}

\def\kms{km s$^{-1}$}

\def\deg{$^{\circ}\,$}
\def\solm{M$_{\odot}\,$}

\def\kms{km s$^{-1}$}

\def\mass{$10^{11}$ M$_{\odot}\,$}

\def\lmass{$10^{10}$ M$_{\odot}\,$}

\def\casgm20{CAS-G-M$_{20}\,$}
\def\m20{M$_{20}\,$}
\begin{document}

\date{Accepted ; Received ; in original form}

\pagerange{\pageref{firstpage}--\pageref{lastpage}} \pubyear{2002}

\maketitle

\label{firstpage}

\begin{abstract}

Utilizing deep Hubble Space Telescope imaging from the two largest
field galaxy surveys, the Extended Groth Strip (EGS) and the COSMOS
survey, we examine the structural properties, and derive the merger 
history for $21,902$ galaxies with M$_{*} > 10^{10}$ \solm at $z < 1.2$.  
We examine the structural CAS parameters of these galaxies, deriving
merger fractions, at $0.2 < z < 1.2$, based on the asymmetry and clumpiness 
values of these systems.   We find that the
merger fraction between $z = 0.2$ and $z = 1.2$ increases 
from roughly $f_{m} = 0.04\pm0.01$ to $f_{m} = 0.13\pm0.01$.  We furthermore
detect, at a high significance, an abrupt drop in the merger fraction at
$z < 0.7$, which appears relatively constant from $z = 0.7$ to $z = 1.2$. We 
explore several fitting formalisms for parameterising the merger
fraction, and compare our results to other structural studies and
pair methods within the DEEP2, VVDS, and COSMOS fields.  We also
examine the basic features of these galaxies, including our selection
for mergers, and the inherent error budget and systematics
associated with finding mergers through structure. We find that
for galaxies selected by M$_{*} > 10^{10}$ \solm, the merger 
fraction can be parameterised by $f_{m} = f_{0} \times (1+z)^{m}$
with the power-law slope $m = 2.3\pm0.4$.  By using the best available 
$z = 0$ prior the slope increases to $m = 3.8\pm0.2$, showing how critical 
the measurement of local merger properties are for deriving the evolution
of the merger fraction.  We furthermore show that the merger fraction
derived through structure is roughly a factor of 3-6 higher than 
pair fractions. Based on
the latest cosmological simulations of mergers we show that this
ratio is predicted, and that both methods are likely tracing the merger
fraction and rate properly.  We calculate, utilising merger time scales
from simulations, and previously published merger fractions within
the Hubble Deep and Ultra Deep Fields, that the merger rate of galaxies with 
M$_{*} > 10^{10}$ \solm
increases linearly between $z = 0.7$ and $z = 3$. Finally, we show
that a typical galaxy with a 
stellar mass of M$_{*} > 10^{10}$ \solm  undergoes between 1-2 major 
mergers at  $z < 1.2$.

\end{abstract}


\section{Introduction}

Unlike the case for stars, the relative size and separation between galaxies is
small, suggesting that galaxy interactions and mergers could be
a major process in driving galaxy formation.  Observational attempts
to determine the role of galaxy mergers in the local universe have
been limited to studies of how star formation and black hole growth
are induced by the merger process (e.g.,  Barton et al. 2000; Ellison
et al. 2008).  It
is however clear that ongoing major galaxy merging is rare in the nearby 
universe, and is likely not a dominant mode for the future evolution of 
galaxies (e.g., Patton et al. 2000).    If galaxy mergers play a major
role in galaxy formation they must have occurred earlier in the 
history of the universe.

Attempts to trace the merger history in the early universe has been
carried out using two primary methods.  The oldest method involves 
measuring the fraction of galaxies which are in
physical pairs at various look-back times, or redshifts (e.g., Patton
et al. 2002; Le Fevre et al. 2000; Lin et al. 2004; Lin et al. 2008;
Bluck et al. 2008).
A newer method, with a considerable background history (e.g., Holmberg
1941; Vorontsov-Velyaminov 1959; Arp 1966), uses the structures of 
galaxies to determine the
merger history (e.g., Conselice 2003; Conselice et al. 2003; Lotz
et al. 2008).    It is, however, not yet known if the two methods
are revealing the same merger history for galaxies, although at
$z \sim 0$, the agreement between the two methods appears good
(e.g., De Propris et al. 2007).

The merger history of galaxies is still largely only beginning to 
be measured with some certainty.  Two issues that still need to be
addressed are the reliability of the current methods for determining 
merger histories, and the relatively small areas thus far used to find
mergers.  Both of these issues limit our ability to accurately
measure the merger history. The reliability argument has been addressed
in great detail in papers such as Conselice (2003), Conselice
et al. (2003), Conselice et al. (2005), and Lotz et al. (2004). Briefly,
the merger history can only be measured using special techniques and
data, such as complete redshift surveys and/or deep Hubble Space
Telescope (HST) imaging.    It turns out that when examining galaxies
that fit a structural merger criteria, nearly all are ongoing major
mergers, as judged by eye and through kinematics (e.g., Conselice 2000a,b;
Conselice 2003; Conselice et al. 2008). This hold for low and high
redshift galaxies (e.g., Conselice et al. 2003), and is also seen
when examining N-body models of the merging process (Conselice 2006;
Lotz et al. 2008b).

Using structure to determine the merger history requires the use of
HST imaging or adaptive optics (AO).  However, the small fields of view used
in HST and AO surveys make it difficult to
measure the merger history with a high certainty, simply due to
the small number of galaxies that have been examined.  Another issue is
that it is difficult to measure the merger history at higher redshifts
due to the fact that there is very little high resolution deep near-infrared 
imaging, which is needed to directly study the rest-frame
optical structures of $z > 1.2$ galaxies.  This is required, as the rest-frame
optical appearance of a galaxy often differs significantly from the 
rest-frame ultraviolet morphology (e.g., Windhorst et al. 2002; 
Taylor-Mager et al. 2007).  Ideally in the future, using multiple band 
data, it is desirable to measure 
parameters and structures on stellar mass images (e.g., Lanyon-Foster
et al. 2009).

For our merger measurements, we utilise the CAS method (e.g., Conselice 2003) 
for determining the 
presence of galaxy mergers in the galaxy population at $z < 1$.  We first
re-evaluate the use of the asymmetry index (Conselice et al. 2000a),
and the CAS method itself for locating major mergers.  We furthermore
use our derived merger fractions to characterise the major galaxy
merger evolution for M$_{*} > 10^{10}$ \solm galaxies at
$z < 3$.  We discuss various ways in which this merger evolution
can be parameterised, investigating power-law, exponential, and
the combination of the two forms.    We conclude that the use
of a $z = 0$ prior in fitting the merger fraction is a significant
factor in the determination of the power-law slope $m$.  We show
that this value of $m$ can vary between $m = 1.5$ and $m = 4$ depending
on how the merger fraction is fit (cf. Bluck et al. 2008 for ultra
massive galaxies).

In this paper we use the two largest HST Advanced Camera for Surveys (ACS) 
imaging data sets - the
Extended Groth Strip (EGS) and the Cosmic Evolution Survey (COSMOS) to 
determine, using $> 20,000$ massive galaxies, the detailed merger history at 
$z < 1.2$.  This is an important
epoch for measuring the merger history, and there is still considerable
uncertainty regarding the merging history during this epoch, which traces
the last half of the universe. By using ACS F814W band
images we are also tracing the structures of these galaxies in the rest-frame 
optical out to $z \sim 1$ and do not need to consider the sometimes 
considerable morphological $k-$corrections when imaging galaxies in the
ultraviolet (e.g., Taylor-Mager et al. 2007).

Our general conclusion is that the merger fraction increases with
higher redshifts to $z \sim 1.2$. We use the latest model based time-scales for galaxies
in pairs to merge, and for galaxies to remain symmetric during
mergers, to calculate
the number of mergers a typical M$_{*} > 10^{10}$ \solm galaxy
will undergo at $z < 3$, as well as the galaxy merger rate. We
conclude that a typical  M$_{*} > 10^{10}$ \solm galaxy will
undergo 1-2 major mergers at $z < 1$. We also investigate how structural 
merger fractions
compare with pair fractions, finding that at a given redshift, within the 
same population of galaxies, the CAS merger
fraction is 3-6 times higher than the pair fraction.  We 
show that the ratio of the time-scales for CAS mergers and 20 kpc pairs
is nearly the same as the merger-pair fraction ratio. Both 
methods therefore appear to trace the same merging population at different 
phases. This is further evidence that
we are indeed measuring correctly the merging properties of
galaxies.

This paper is organised as follows: \S 2 includes a discussion of the 
data sources we use in this paper, and how we select our sample, including a 
description of our 
morphological and structural analyses, and the stellar masses we utilise, 
\S 3 is a discussion of our
results, including a description of the merger history up to $z \sim 1.2$, and \S 4 is our 
summary and conclusions.   We use a standard cosmology
of H$_{0} = 70$ km s$^{-1}$ Mpc$^{-1}$, and 
$\Omega_{\rm m} = 1 - \Omega_{\lambda}$ = 0.3 throughout.

\section{Data and Sample Selection}

\subsection{Data}

The data we use in the paper originate from the Extended Groth
Strip (EGS) survey (Davis et al. 2007), and the COSMOS survey
(Scoville et al. 2007).  The selection of galaxies which
we analyse is simply those 
systems which have a stellar mass M$_{*}$ $> 10^{10}$ \solm. As the EGS
and the COSMOS fields have different data sets, and methods for 
galaxy detection, and for
measuring stellar masses, we consider both data sets separately
in what follows. In total there are 21,902 M$_{*}$ $> 10^{10}$ \solm
galaxies at $z < 1.2$ in our sample, with 2,388 galaxies in the EGS and
19,514 in the COSMOS field.

The main data set for this paper is the COSMOS field, which is
by far the largest mosaic of Hubble Space Telescope imaging using
the Advanced Camera for Surveys (ACS).  The COSMOS 
ACS coverage is 1.8 deg$^{2}$ and is imaged in the F814W (I) band
over this entire area with 1 orbit depth per pointing, for
a total of 590 orbits.  The 50\% completeness of the COSMOS
ACS data is I$_{\rm AB} = 26$ (Scoville et al. 2007).  The
COSMOS field also has extensive optical data, described
in Mobasher et al. (2007), from which stellar masses and
photometric redshifts are computed.  The photometric data
in which these quantities are derived include optical data
in the $u^{*}$ band from the CFHT, $BgVriz$ imaging from
the SuprimeCam on Subaru, $i^{-}$ imaging from CHFT and
$K_{S}$ imaging from FLAMINGOS taken at CTIO and Kitt
Peak (Mobasher et al. 2007).  The I$_{814}$ ACS imaging is also
included in the analysis.  The seeing for this optical and NIR ground
based imaging is roughly 1$\arcsec$, with depths that
range from 21.5 AB in $K_{S}$ to $\sim 27$ in the $r$-band.
The depth in every band is in any case well matched to image the most
massive galaxies at $z < 1.2$ at a high S/N.

The ACS imaging of the EGS field  covers a 10.1\arcmin $\times$ 
70.5\arcmin\,
strip, for a total area of 0.2 deg$^{2}$.  This ACS imaging
is discussed in Lotz et al. (2008) and Conselice et al. (2008b), 
and is briefly described
here.  The imaging consists of 63 tiles imaged in both the
F606W (V) and F814W (I) bands.  The 5-$\sigma$ depths reached in these
images are V = 26.23 (AB) and I = 27.52 (AB) for point
sources, and about two magnitudes brighter for extended objects.

Our sample for the EGS comes from those systems which have K-band
data taken as part of the Palomar Observatory Wide-field Infrared
Survey (POWIR; Bundy et al. 2006; Conselice et al. 2007a,b; 2008b). The
POWIR survey was designed to obtain deep K-band and J-band data over a 
significant ($\sim$1.5 deg$^2$) area.   Observations were 
carried out between September 2002 and October 2005 over a total of $\sim
70$ nights. This survey
covers the GOODS field North (Giavalisco et al. 2004; Bundy et al. 2005),
the Extended Groth Strip (Davis et al. 2007), and three other fields 
that the DEEP2 team
has observed with the DEIMOS spectrograph (Davis et al. 2003).  The total
area we cover in the K-band is 5524 arcmin$^{2}$ = 1.53 deg$^{2}$, with
half of this area imaged in the J-band.  Each of our fields reach  5-$\sigma$
depths between K$_{\rm s,vega} = 20.5 - 21.5$ for point sources, as 
measured in a 2\arcsec\, diameter aperture with the EGS.

Our K$_{\rm s}$-band data were acquired utilising the WIRC camera 
on the Palomar 5 meter telescope.  WIRC has an effective field of view of 
$8.1\arcmin \times 8.1\arcmin$, with a pixel scale
of 0.25\arcsec pixel$^{-1}$.   Our K$_{\rm s}$-band data were taken using
30 second integrations, with four exposures per pointing, while the J-band 
observations were taken with 120 second exposures per pointing. Typical total 
exposure times are one and two hours for both bands. Our reduction procedure 
follows
standard methods for combining NIR ground-based imaging, and is described
in more detail in Bundy et al. (2006). The resulting seeing FWHM in the 
K$_{\rm s}$-band imaging ranges from 0.8\arcsec to 1.2\arcsec, and is 
typically 1.0\arcsec (e.g., Bundy et al. 2006).  
Other data sets within the EGS we use consist of: optical
imaging from the CFHT over all fields, imaging from the Advanced Camera for 
Surveys (ACS) on Hubble, and spectroscopy from the 
DEIMOS spectrograph on the Keck II telescope (Davis et al. 2003).  A summary
of these ancillary data sets are included in Davis et al. (2007) and
Conselice et al. (2007b).

The optical imaging of the EGS is taken with the CFHT 3.6-m telescope.
This optical data consists
of imaging in the B, R and I bands taken with the CFH12K camera - a 12,288 
$\times$ 8,192 pixel CCD mosaic with a pixel scale of 0.21\arcsec.  
The integration times for these observations are 1 hour in $B$ and $R$, and
2 hours in $I$, per pointing, with a R-band 5-$\sigma$ depth of 
$R_{\rm AB} \sim 25.1$, and similar depths at $B$ and $I$.
The seeing for the optical imaging is
roughly the same as that for the NIR imaging, and we measure photometry
using a 2\arcsec\, diameter aperture.
  
\subsection{Redshifts}

We utilise both spectroscopic and photometric redshifts for the
galaxies we study in both the EGS and COSMOS fields.  The only
field which has extensive available spectroscopy
is the EGS.  EGS Keck spectra were acquired with the DEIMOS spectrograph 
as part of the DEEP2 redshift survey (Davis
et al. 2003).  Target selection for the DEEP2 spectroscopy
was based on the optical properties of the galaxies detected in the
CFHT photometry, with the basic selection criteria $R_{\rm AB} < 24.1$.   
DEEP2 spectroscopy was acquired through
this magnitude limit, with no strong colour cuts applied to the selection.
About 10,000 redshifts are measured for galaxies within the EGS.  The 
sampling rate for galaxies that meet the selection criteria is
60\%.

This DEIMOS spectroscopy was obtained using the
1200 line/mm grating, with a resolution R $\sim 5000$ covering
the wavelength range 6500 - 9100 \AA.  Redshifts were measured through
an automatic method comparing templates to data, and we only utilise
those redshifts measured when two or more lines were
identified, providing very secure measurements.  Roughly 70\% of all
targeted objects result in secure redshifts. 

As the COSMOS field is the major contributor to the data
used in this paper we give a description of the photometeric redshifts
which we utilise.  The details of its photometric redshifts and the
catalog we use is described in great detail in Mobasher et al. (2007).
We give a short summary here.
The COSMOS photometric redshifts are measured through a standard template
fitting technique. The templates used by Mobasher et al. (2007) are
galaxies of various spectral types from ellipticals to starbursts
across optical rest-frame wavelengths.   The photometric redshifts we use
were tested by comparing with 868 spectroscopic redshifts from the
zCOSMOS survey, where the rms scatter in the agreement
between photometric redshifts and spectroscopic redshifts is
$\sigma(\Delta(z))$ = 0.031, where $\Delta (z) = 
(z_{\rm phot} - z_{\rm spec})/(1+z_{\rm spec})$ (Mobasher
et al. 2007).  Less than 2.5\% of the galaxies are outliers
in the agreement between spectroscopic and photometric redshifts.  Mobasher
et al. (2007) furthermore test their method of measuring photometric
redshifts by calculating values using three different codes, and
find a good agreement with their own calculation.

We utilise our own photometric redshifts within the EGS (e.g., 
Bundy et al. 2006; Conselice et al. 2007b, 2008b).  The determination of
the EGS photometric redshifts is done in a different way than in the
COSMOS field.   Within the EGS, photometric redshifts are based on
the optical+near infrared imaging, in the BRIJK  bands, and are fit 
in two ways, 
depending on the brightness of a galaxy in the optical.   For galaxies that
meet the spectroscopic criteria, $R_{\rm AB} < 24.1$, we utilise a neural
network photometric redshift technique to take advantage of the
vast number of secure redshifts with similar photometric data.  Most
of the $R_{\rm AB} < 24.1$ sources not targeted for spectroscopy should be 
within our redshift range of interest, at $z < 1.4$.    

The neural network 
fitting is done through 
the use of the ANNz (Collister \& Lahav 2004) method and code.
To train the code, we use the $\sim 5000$ redshifts in the EGS, which
has galaxies spanning our entire redshift range.  The training of the 
photometric
redshift fitting was in fact only done using the EGS field, whose
galaxies are nearly completely selected based on a magnitude
limit of $R_{\rm AB} < 24.1$. We then use this training to calculate the
photometric redshifts for galaxies with $R_{\rm AB} < 24.1$.   
The agreement between our photometric redshifts and our ANNz 
spectroscopic redshifts is very good 
using this technique, with $\delta z/(1+z) = 0.07$ out
to $z \sim 1.4$. The agreement is even better for the M$_{*} >$ \mass
galaxies where we find $\delta z/(1+z) = 0.025$ across all of our
four fields (Conselice et al. 2007b).   

For galaxies which are fainter than $R_{\rm AB} = 24.1$ in the EGS
we utilise photometric redshifts using Bayesian techniques, and the 
software from Benitez (2000).  For an object to have a photometric redshift
we require that it be detected at the 3-$\sigma$ level in all
optical and near-infrared (BRIJK) bands, which in the R-band
reaches $R_{\rm AB} \sim 25.1$. We 
optimised our results, and correct for systematics, through 
the comparison with spectroscopic redshifts, resulting
in a redshift accuracy of $\delta z/z = 0.17$ for $R_{\rm AB} > 24.1$ systems. 
These $R_{\rm AB} > 24.1$ galaxies are, however, only a very small part of our
sample.  Furthermore, all of these systems are at $z > 1$.  

\subsection{Stellar Masses}

Since stellar masses are a critical aspect of this analysis,
we go into some detail for how these are calculated.
The stellar masses for the COSMOS survey are  taken from
Mobasher et al. (2007).  These stellar masses are computed
through the use of K-band imaging and the measured rest-frame
$(B-V)_{0}$ colours of galaxies within COSMOS.  Mobasher
et al. (2007) use this measured colour to obtain a mass to
light ratio through the relation:

\begin{equation}
M/L_{v} = -0.628 + 1.305 (B-V)_{0}
\end{equation}

\noindent which assumes a Salpeter IMF within a mass range
of 0.1 \solm to 100 \solm.  The colours used in this analysis are
not directly measured from the data, but are taken from the
best fit templates of various types (E, Sa, Sb, Sc, Im and starburst)
(see Mobasher et al. 2007).  Due to limited K-band data over
the COSMOS field, the stellar masses of these galaxies are
calculated via rest-frame V-band magnitudes with
the equation:

\begin{equation}
{\rm log} (M_{\rm stellar}/M_{\odot}) = M/L_{V} - 0.4 *(M_{V} - 4.82).
\end{equation}

\noindent Mobasher et al. (2007) discuss using V-band luminosities for
measuring stellar masses rather than K-band, which is very
shallow over the COSMOS fields.    They find for a subset of their
sources, which are detected in the K-band, an agreement with the
V-band based stellar masses, these are however the reddest, and thus 
perhaps the most evolved galaxies, and therefore may not be representative.

Within the EGS field we match our K-band selected catalogs to the CFHT 
optical data to obtain spectral energy distributions (SEDs) for all of our sources, 
resulting in measured BRIJK magnitudes.   From these we compute stellar masses
 based on  the methods and results
outlined in Bundy, Ellis, Conselice (2005) and Bundy et al. (2006). All our stellar masses
are furthermore normalised by the observed rest-frame K-band light, which
is roughly at rest-frame $\sim 1\,\mu m$, or redder, for most galaxies. We
also convert these stellar masses, which are calculated in equation (2)
using a Salpeter IMF to a Chabrier IMF.

The basic mass fitting method within the EGS consists of fitting a grid of 
model SEDs constructed
from Bruzual \& Charlot (2003) (BC03) stellar population synthesis models, with
different star formation histories. We use an exponentially declining model
to characterise the star formation history, with various ages, 
metallicities and dust contents included.  These models are parameterised
by an age, and an e-folding time for parameterising the star formation 
history, where SFR $\sim\, e^{\frac{t}{\tau}}$.  The values of $\tau$ are
randomly selected from a range between 0.01 and 10 Gyr, while the age
of the onset of star formation ranges from 0 to 10 Gyr. The metallicity
ranges from 0.0001 to 0.05 (BC03), and the dust content is parametrised
by $\tau_{\rm v}$, the effective V-band optical depth for which we use values
$\tau_{\rm v} = 0.0, 0.5, 1, 2$.     Although we vary several parameters,
the resulting stellar masses from our fits do not depend strongly on the
various selection criteria used to characterise the age and the metallicity
of the stellar population.

It is important to realise that these  parameterisations are
fairly simple, and it remains possible that stellar mass from
older stars is missed under brighter, younger, populations. While
the majority of our systems are passively evolving older stellar populations,
it is possible that up to a factor of two in stellar mass is missed in any 
star bursting blue systems.  However, stellar masses measured through
our technique are roughly the expected factor of 5-10 smaller than
dynamical masses at $z \sim 1$ using a sample of disk galaxies
(Conselice et al. 2005b), demonstrating their inherent reliability.

We match magnitudes derived from these model 
star formation histories to the actual data to obtain a measurement
of stellar mass using a Bayesian approach.
We calculate 
the likely stellar mass, age, and absolute magnitudes for each galaxy at all 
star formation histories, and determine stellar masses based on this
distribution.  Distributions with larger ranges of stellar masses
have larger resulting uncertainties. While parameters such 
as age, e-folding time, metallicity, etc. are not likely accurately fit
through these
calculations due to various degeneracies, the stellar mass is robust.  
Typical errors for our stellar masses are 0.2 dex from the width
of the probability distributions.  There are also uncertainties from
the choice of the IMF.  Our stellar masses utilise the
Chabrier (2003) IMF. There are additional
random uncertainties due to photometric errors.  The resulting
stellar masses thus have a total random error of 0.2-0.3 dex,
roughly a factor of two.  

Furthermore, there is the issue of whether or not our stellar masses
are overestimated based on using the Bruzual \& Charlot (2003)
models.  It has recently been argued by Maraston (2007), among
others, that a refined treatment of thermal-pulsating AGB stars 
in the BC03 models result in stellar masses that can be
too high by a factor of a few. While we consider an uncertainty
of a factor of two in our stellar masses, it is worth investigating
whether or not our sample is in the regime where the effects of
a different treatment of TP-AGB stars in e.g., Maraston (2007) will 
influence our mass measurements.  This has been investigated in 
Maraston (2005) who have concluded that galaxy stellar masses computed
with an improved treatment of TP-AGB stars are roughly 
50-60\% lower.

However, the effect
of TP-AGB stars is less important at our rest-frame wavelengths probed 
than at longer wavelengths, especially in the rest-frame IR. The EGS survey 
is K-selected, and the
observed K-band is used as the flux in which stellar masses are computed. The
rest-frame wavelength probed with the observed K-band ranges from 0.7$\mu$m
to 1.5$\mu$m where the effects of TP-AGB stars are minimised.
The ages of our galaxies are also older than the ages where
TP-AGB stars have their most effect (Maraston 2005).  
To test this, after our analysis was finished, we utilised
the newer Bruzual and Charlot (2009, in prep) models, which include a new 
TP-AGB star prescription, on our massive galaxy sample. From this we
find on average a $\sim$ 0.07 dex smaller stellar mass using the newer
models.  At most, the influence of TP-AGB stars will decrease our
stellar masses by 20\%.  The effect of this would decrease
the number of galaxies within our sample, particularly those
close to the M$_{*} =$ \lmass boundary.  This systematic error is however
much smaller than both the stellar mass error we assume (0.3 dex), and
the cosmic variance uncertainties, and thus we conclude it is not
a significant factor for our analysis.

\subsection{The Extended CAS Structural Analysis}

We use the CAS (concentration, asymmetry, clumpiness) parameters to measure
the structures of our $z < 1.2$ galaxies quantitatively.  We include in our
analysis the measurement of the Gini and \m20 parameters (e.g.,
Lotz et al. 2008).  The CAS/Gini/\m20 parameters  are a 
non-parametric method for measuring the forms and structures of galaxies 
in resolved 
CCD images (e.g., Conselice et al. 2000a; Bershady et al. 2000; 
Conselice et al. 2002;  Conselice 2003; Lotz et al. 2008).   The
basic idea behind these parameters is that galaxies have light distributions
that reveal their past and present formation modes (Conselice 2003). 
Furthermore, well-known galaxy types in the nearby universe fall in 
well defined regions of the CAS parameter space.  For example, the 
selection $A > 0.35$ locates systems which are nearly all major
galaxy mergers in the nearby universe (e.g., Conselice et al. 2000b; 
Conselice 2003; Hernandez-Toledo et al. 2005; Conselice 2006b).
In addition to the classic CAS parameters, we also investigate
the use of the similar Gini and M$_{20}$ parameters (Lotz et al. 2008).  
We give a brief description of these parameters below.

The way we measure structural parameters on our Hubble images varies slightly
from what has been done earlier in the Hubble Deep Field, and GOODS 
imaging (e.g., Conselice et al. 2003a; Conselice et al. 2004). The basic
measurement procedure, after cutting out the galaxy into a smaller image, is to
first measure the radius in which the parameters are computed. The
radius we use for all our indices is defined by the Petrosian radii, 
which is the radius where the surface brightness at a given
radius is 20\% of the surface brightness within that radius (e.g.,
Bershady et al. 2000; Conselice 2003).   

We use circular apertures for
our Petrosian radii and quantitative parameter estimation.  We begin
our estimates of the galaxy centre for the radius measurement at
the centroid of the galaxy's light distribution.  Through modelling
and various tests, it can be shown that the resulting radii do not
depending critically on the exact centre, although the CAS and other
parameters do (Conselice et al. 2000; Lotz et al. 2004).  The exact
Petrosian radius we use to measure our parameters is

$$R_{\rm Petr} = 1.5 \times r(\eta = 0.2),$$

\noindent where $r(\eta = 0.2)$ is the radius where the surface
brightness is 20\% of the surface brightness within that radius.

A very important issue, especially for faint galaxies, is how to account 
for background light and noise. For faint galaxies there
is a considerable amount of noise added due to the sky, which must
be corrected.  Through various test, outlined in detail
in Conselice et al. (2009, in prep), we conclude that the proper
way to correct parameters for the background requires that the selected
background area be close to the object of interest. This is only an issue
for faint galaxies, and for galaxies imaged on large mosaics which
have a non-uniform weight map, and whose noise characteristics vary
across the field.  By using a background near each object we alleviate these
issues as the noise properties do not vary significantly over 
$\sim 0.5 - 1$ arcmin, where the galaxy and the
background area are selected.  We review below how the CAS and Gini/\m20
parameters are measured. For more detail see Bershady et al. (2000),
Conselice et al. (2000), Conselice (2003), and Lotz et al. (2008).

\subsubsection{Asymmetry}

The asymmetry of a galaxy is measured by taking an original galaxy 
image and rotating it 180 degrees about its centre, and then
subtracting the two images (Conselice 1997). There are corrections done for
background, and radius (explained in detail in Conselice et al. 2000a).
Most importantly, the centre for rotation is decided by an iterative
process which finds the location of the minimum asymmetry.  The formula
for calculating the asymmetry is given by:

\begin{equation}
A = {\rm min} \left(\frac{\Sigma|I_{0}-I_{180}|}{\Sigma|I_{0}|}\right) - {\rm min} \left(\frac{\Sigma|B_{0}-B_{180}|}{\Sigma|I_{0}|}\right)
\end{equation}

\noindent Where $I_{0}$ is the original image pixels, $I_{180}$ is the image
after rotating by 180\deg.  The background subtraction using light from a
blank sky area, called $B_{0}$, are critical for this process, and must 
be minimised in the same way as the original galaxy itself.  A lower value 
of $A$ means that a galaxy has a 
higher degree of rotational symmetry which tends to be found in
elliptical galaxies.
Higher values of $A$ indicate an asymmetric light distribution, which are 
usually found in spiral galaxies,  or in the more extreme case, merger 
candidates. 

\subsubsection{Concentration}

Concentration is a measure of the intensity of light contained within a 
central region in comparison to a larger region in the outer-parts of a
galaxy.  The exact definition is the ratio of two circular radii which 
contain 20\% and 80\% ($r_{20}$, $r_{80}$) of the total galaxy flux,

\begin{equation}
C = 5 \times {\rm log} \left(\frac{r_{80}}{r_{20}}\right).
\end{equation}

\noindent This index is sometimes called C$_{28}$.  
A higher value of $C$ indicates that a larger amount of light 
in a galaxy is contained within a central region.   This
particular measurement of the concentration correlates well with 
the mass and halo properties of galaxies in the nearby
universe (e.g., Bershady et al. 2000;  Conselice 2003).

\subsubsection{Clumpiness}

The clumpiness ($S$) parameter is used to 
describe 
the fraction of light in a galaxy which is contained in clumpy distributions.
Clumpy galaxies have a relatively large amount of
light at high spatial frequencies, 
whereas smooth systems, such as elliptical galaxies contain light at low 
spatial frequencies. Galaxies which are undergoing star formation tend to 
have very clumpy structures, and high $S$ values.  Clumpiness can be 
measured in a number of ways, the most common method used,
as described in Conselice (2003) is,

\begin{equation}
S = 10 \times \left[\left(\frac{\Sigma (I_{x,y}-I^{\sigma}_{x,y})}{\Sigma I_{x,y} }\right) - \left(\frac{\Sigma (B_{x,y}-B^{\sigma}_{x,y})}{\Sigma I_{x,y}}\right) \right],
\end{equation}

\noindent where, the original image $I_{x,y}$ is blurred to produce 
a secondary image,  $I^{\sigma}_{x,y}$.  This blurred image is
then subtracted from the original image leaving a 
residual map, containing only high frequency structures in
the galaxy (Conselice 2003). To quantify this, we normalise the
summation of these residuals by the original galaxy's total light, and
subtract from this the residual amount of sky after smoothing
and subtracting it in the same way.  The size of the smoothing kernel 
$\sigma$ is
determined by the radius of the galaxy, and is $\sigma = 0.2 \cdot 1.5
\times r(\eta = 0.2)$ (Conselice 2003).  Note that the centres of galaxies are
removed when this procedure is carried out.  

\subsubsection{Gini and \m20 Coefficients}

The Gini coefficient is a statistical tool originally used in economics to 
determine the distribution of wealth within a population, with higher values 
indicating a very unequal distribution (Gini of 1 would mean all wealth/light 
is in one person/pixel), while a lower value indicates it is distributed 
more evenly amongst the population (Gini of 0 would mean everyone/every pixel 
has an equal share).   The value of G is defined by the Lorentz curve 
of the galaxy's light distribution, which does not take into 
consideration spatial position.  Each pixel is ordered by its
brightness and counted as part of the cumulative distribution (see
Lotz et al. 2004, 2008).

The \m20 parameter is a similar parameter to the concentration in that it 
gives a value that indicates 
whether light is concentrated within an image; it is however calculated 
slightly differently.  The total moment of light is calculated by summing the 
flux of each pixel multiplied by the square of its distance from the centre.  
The centre is deemed to be the location where \m20 is minimised
(Lotz et al 2004).  The value of \m20 is the moment of the 
fluxes of the brightest 20\% of light in a galaxy, which is
then normalised by the total light moment for all pixels (Lotz
et al. 2004, 2008).

\section{Analysis}

\subsection{Characteristics of the Sample}

This analysis is unique in several ways. First, it is based
on two major data sets - the COSMOS and the EGS surveys. The total
number of galaxies we study for our morphological analysis is also
by far the largest ever published to date.  Similar previous work using the 
GOODS fields, the Hubble Deep Fields, and the Hubble
Ultra Deep Field (e.g.,  Conselice et al. 2003;
Conselice et al. 2003, 2004; Conselice et al. 2005; Lotz et al. 2008)
used smaller samples, and thus were able to check systematics in a way
we cannot due to our large sample of galaxies.  The large number of
galaxies in our sample makes the type of checking done in previous
work largely impossible (cf. Jogee et al. 2008 for this type of
analysis done on the GEMS data set).  

The sample of galaxies in which we use from the EGS is described in
great detail in Conselice et al. (2007b), and we refer all readers to
this paper for characteristics of the sample.  The sample we use
for the COSMOS field is however not as well characterised, nor as
well understood, or as accurately calibrated.    We can however, investigate
some basic features of the COSMOS data set to determine how it compares
with the better calibrated (for our purposes) EGS data set and sample,
described in Conselice et al. (2007b).  
 
Figure~1 shows how 
stellar masses are distributed with redshift in the COSMOS field,
demonstrating some clustering features which may bias the results,
if the merger history is a strong function of environment. The COSMOS
field is known to contain a mixture of environments from single
galaxies to large groups (Scoville et al. 2007), and our sample is
taken from all of these environments.

\begin{figure}
 \vbox to 120mm{
\includegraphics[angle=0, width=90mm]{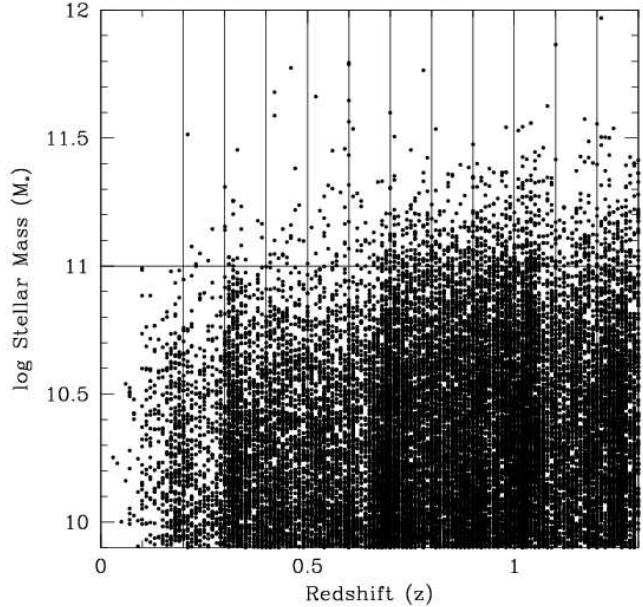}
 \caption{The stellar mass vs. redshift ($z$) relation for the COSMOS
field.  The vertical lines show the redshift limits for the various
bins we use within our analysis.  The horizontal line represents the
cut at log M$_{*} = 11$ we use in some of our analysis.  Note that
the clustered redshift peaks are due to real over-densities.  The distribution
within the EGS for the same stellar mass range is similar (Conselice
et al. 2007a).}
} \label{sample-figure}
\end{figure}

The other feature through which we investigate the sample is the
CAS diagrams of concentration-asymmetry and asymmetry-clumpiness (Conselice
2003).  These diagrams reveal which type of galaxies are likely to be within
the sample.   What we found in our initial analysis is that
a significant number of systems within the COSMOS field 
had concentration values near $C = 2$, which is typically found for
stars.  Visual inspection of these images reveals that this is indeed
the case - many stars appear to be within the released COSMOS galaxy
catalogs. On the
other hand, our star/galaxy separation in the EGS (see Conselice
et al. 2007 and references therein) was very effective at removing
stars and no such contamination is found in the $C-A$ diagrams.  
We utilize the fact that
stars have a well defined concentration and small measured radii to
remove these systems from our COSMOS CAS catalog.

Figure~2 shows the resulting concentration-asymmetry diagram at
$z > 0.75$ and $z < 0.75$ for the combined COSMOS and EGS sample.  
What this reveals is that there is a wide diversity
of structures for galaxies at all redshifts, as shown using
smaller samples in e.g., Conselice et al. (2003, 2005) and
Lotz et al. (2008).  This includes galaxies which would be
classified at $z = 0$ as early types, late-types, and systems
which are asymmetric enough to be ongoing major mergers
(Conselice 2006a).  

One of the most interesting things concerning Figure~2 is the clear
bimodality of the structural parameters for these systems.
At both redshifts bins, there is a peak in the density contours
near the mid/early-type range, and another peak well within
the late-type range. This bimodality is present within
all redshifts, and is the same bimodality found using
similar structural parameters by Zamojski et al. (2007) 
within COSMOS.  At $z < 0.75$ we find one peak at
$C = 3.9$ and $A = 0.1$, and the other peak at
$C = 2.9$ and $A = 0.2$.  These two values of $C$ and
$A$ are equivalent to early-types and late-type galaxies,
and is likely similar to the bimodality found within the
colour-magnitude diagram (e.g., the red-sequence and
blue cloud).

Another interesting feature of Figure~2 is that the peak
location for the centres of the bimodality differs slightly
with redshift, becoming less concentrated and
more asymmetric at high redshift. This change is
$\delta A$ = 0.05 and $\delta C$ = 0.2. The peaks of
the bimodality are also softer, suggesting a larger
scatter about the bimodality seen at lower redshifts.
The slightly lower values are in principle due solely
to redshift effects, or morphological k-corrections,
both of which are roughly of the order of the changes
seen here.

\subsection{Selection of Mergers}

Before we can effectively study the merger history for our
sample, we must select a robust sample of mergers, as cleanly
as possible (e.g., Conselice et al. 2008a). There are
several ways to do this.  The classical rest-frame optical
CAS definition for determining whether a system is undergoing a 
merger is given by (Conselice 2003):

\begin{equation}
A > 0.35\,{\rm and}\, A > S,
\end{equation}

\noindent That is, the asymmetry $A$ must be larger than 0.35 and
the asymmetry must exceed the value of the clumpiness of the
galaxy.  This selection will nearly always cleanly find galaxies
in mergers as revealed through nearby samples of galaxies
(Conselice 2003; De Propris et al. 2007), and through N-body simulations
of the merger process (Conselice 2006).
Tests of how well this criteria does at higher redshifts are limited
to either small samples (Conselice et al. 2008a), and to the GEMS
survey which was based on F606W (V) band imaging of galaxies (Jogee et 
al. 2008).  These
studies however reveal a few features that we can use to determine
how well the criteria given in equation (6) does for finding galaxies
undergoing mergers.

The first is that although the Jogee et al. (2008) study of GEMS
uses the F606W band, at $z = 0.4$, it samples rest-frame optical light, 
as we also do for nearly all our galaxy
sample.  Since Jogee et al. (2008) classify all their galaxies
by eye (which we do not), and use the exact same methodology and
CAS code as we do, we can use their analysis to determine what
fraction of galaxies we are finding using criteria from eq. (6) which
are actual merging galaxies.  Jogee et al. (2008) find that at their
lowest redshift, the contamination by non-merging galaxies, as defined
by visual inspection, is $\sim 30$\%. A significant fraction of
this 30\% however are irregular galaxies, some of which could be
systems in some phase of a merger.  Conselice et al. (2008a) within the UDF,
found that over the entire redshift range of $0.4 < z < 3$ the contamination
is lower, roughly 15\%. These differences are likely due to the 
difficultly of defining what a merger is based on visual appearances.

We also use the clumpiness-asymmetry diagrams of our combined
sample and the concentration-asymmetry diagram discussed in \S 3.1
to get an idea for how well the merger criteria selects merging
systems. First, we note that in Paper I (Conselice et al. 2008a) 
the clumpiness-asymmetry relation found for normal galaxies in
Conselice (2003) does not do a good job in tracing the relation between
asymmetry and clumpiness for normal, non-merging, galaxies.  For
systems at $z > 0.4$ the clumpiness values are lower than their
corresponding asymmetry.  This is due to either to the structures
of galaxies being actually less clumpy, or in the way the
clumpiness is defined by Conselice (2003), or because of decreased
resolution and S/N, the clumpiness decreases.

We reevaluate the criteria for how the asymmetry and clumpiness
values for normal galaxies change for galaxies at $z > 0.4$ through
the use of the Hubble Ultra Deep Field (described in Paper I).
The relation between $S$ and $A$ for all galaxies in the Conselice
et al. (2008a) sample is shown in Figure~3.  The solid line in 
Figure~3 is the best fit relation between the asymmetry and
clumpiness indices for the normal galaxies - i.e., those that are not
merging or peculiar at $z > 0.2$.  This relation is given by

\begin{equation}
A = (0.99\pm0.14) \times S + (0.05\pm0.01),
\end{equation}

\noindent which is roughly $A \sim S$.  
In Figure~4 we show the relation between $A$ and $S$ for the 
COSMOS+EGS sample. As can be seen, there is a general correlation between these
two parameters, with a single peak, and a distribution around 
$S = 0.1$ and $A = 0.15$.  Figure~4 also shows that the average 
asymmetry of these galaxies slightly increases at higher redshifts, with
the clumpiness index remaining roughly the same at the peak value.




\begin{figure*}
 \vbox to 120mm{
\includegraphics[angle=0, width=180mm]{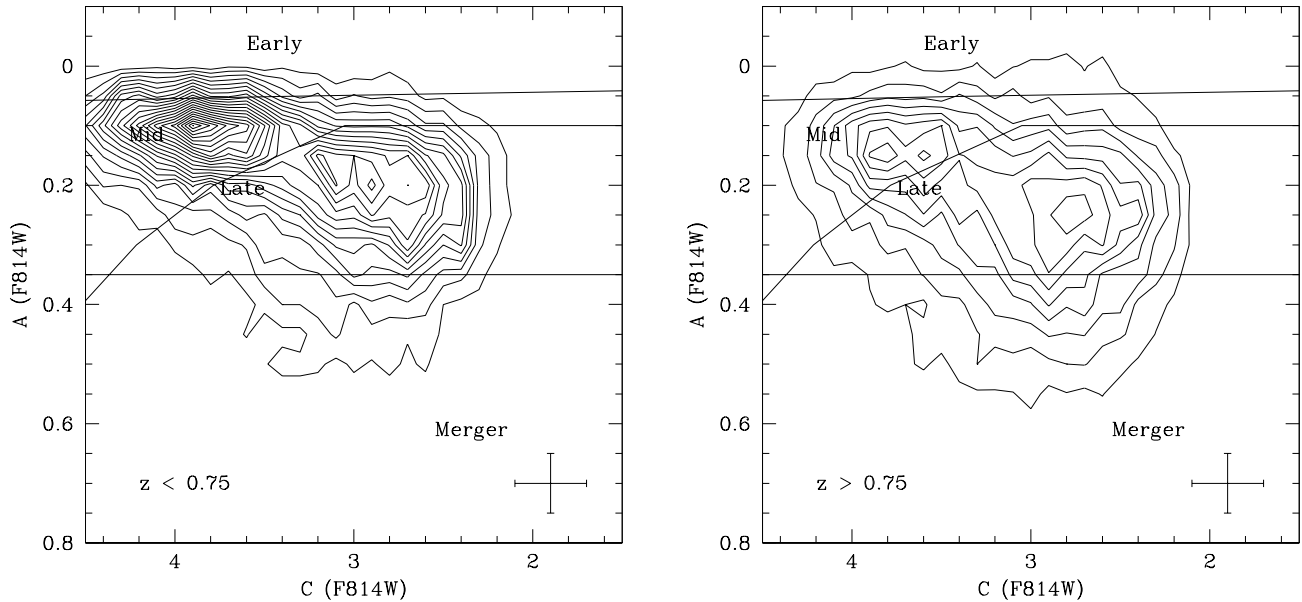}
 \caption{The relationship between the asymmetry ($A$) and the
concentration ($C$) for galaxies within the combined EGS+COSMOS
sample we examine in this paper.  Due to the large number of
points we have used contours to examine this distribution.  The solid
lines denote regions for galaxies of different types which 
are labelled within their respective regions. In general, galaxies
which are at $A > 0.35$ are those which are found to
be ongoing major mergers.  We divide our sample into two roughly
equivalent sample sizes, with the galaxies at $z < 0.75$ shown
on the left and those at $z > 0.75$ plotted on the right. As
can be seen, and discussed within the paper, each redshift
bin displays a bimodality in the galaxy population. The higher
redshift bins shows peaks in this bimodality that are slightly
offset from the lower redshift bin, but the galaxies at $z > 0.75$
also show a large distribution of points, with the peaks not as
well defined. }
\vspace{4cm}
} \label{sample-figure}
\end{figure*}

\begin{figure}
 \vbox to 140mm{
\includegraphics[angle=0, width=90mm]{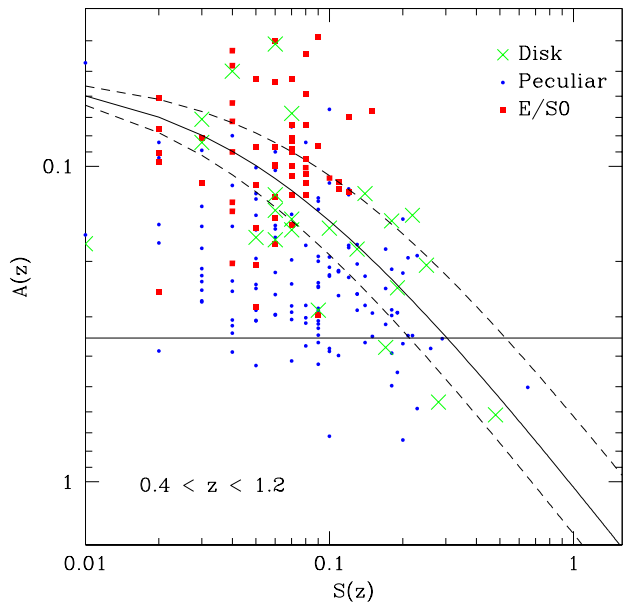}
 \caption{The relation between asymmetry and clumpiness
for galaxies in the Hubble Ultra-Deep Field at $0.4 < z < 1.2$ 
from paper I (Conselice et al. 2008a). Shown are the three main
types of galaxies classified within the UDF (ellipticals, disks
and peculiars).  The solid line shows the best fit relation between
the $A$ and the $S$ parameters for the normal galaxies (disks
and ellipticals). The dashed line shows the 3-$\sigma$ deviation for
this fit.  As can be seen, the disk galaxies which have an asymmetry
$A > 0.35$ generally follow this relation and have a relatively
high asymmetry value for their clumpiness. }
} \label{sample-figure}
\end{figure}

\begin{figure*}
 \vbox to 120mm{
\includegraphics[angle=0, width=180mm]{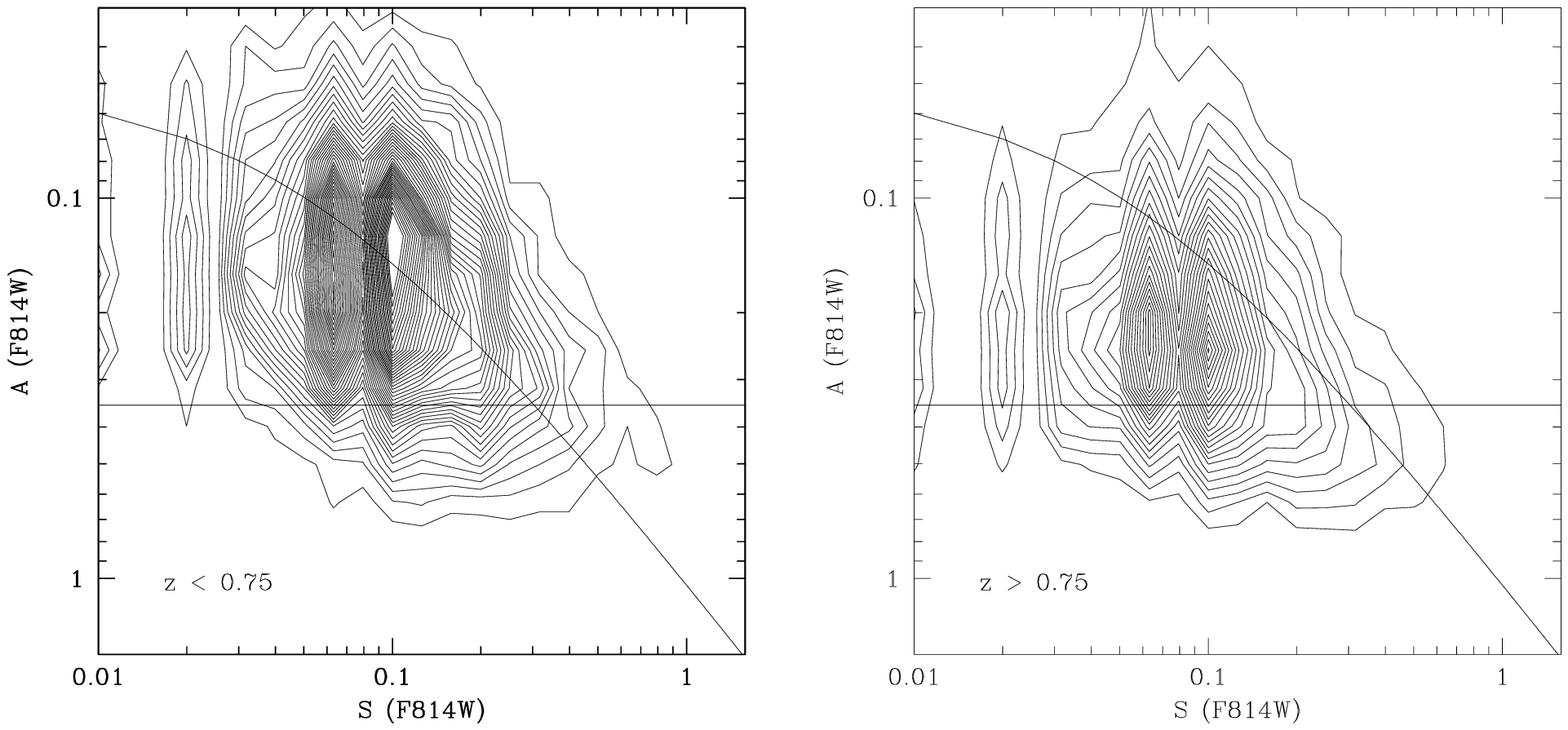}
 \caption{The relationship between the asymmetry ($A$) and the
clumpiness ($S$) indices for galaxies within our combined EGS+COSMOS
sample. The galaxies in this figure and the division between
redshifts is the same as in Figure~2. The horizontal solid line shows the 
$A = 0.35$ limit for finding mergers, and the curved solid line shows
the relation between $A$ and $S$ for normal galaxies at $z < 1.2$ as
found in the UDF and plotted in Figure~3.}
\vspace{-3cm}
} \label{sample-figure}
\end{figure*}

\subsection{The Merger Fraction}

\subsubsection{Merger Fractions}

Understanding the merger history of galaxies is important for
deriving how the merger process drives the assembly of galaxies,
as well as its possible role in the triggering of star formation
and AGN activity.
The merger fraction history at $z > 0.2$ is however considered somewhat
controversial, due to published values of the derived merger  history
differing (e.g., Conselice et al. 2003a; Lotz et al. 2008; de Ravel
et al. 2008).  In this
section we examine the merger history for massive galaxies 
(M$_{*} > 10^{10}$ \solm) using our combined EGS and COSMOS data sets.

One of the benefits of using the CAS system for finding mergers is that 
it allows us to quantify the merger fraction, merger rates, and thus the 
number of mergers occurring in a galaxy population  (Conselice et al. 2003a; 
Conselice 2006b).  The reason for this is that the CAS method has been
well calibrated in terms of the types of galaxies picked up with various
selection criteria at different redshifts, as well as estimates for
the time-scale sensitivity and the ability to pick up mergers with
different mass ratios.   In this section we investigate the merger fractions 
based on standard techniques we have developed (e.g., Conselice 2003; 
Conselice 2006; Bridge et al. 2007; \S 3.2).

There are a few caveats
to measuring the merger fraction which we must consider before using these 
values to determine how galaxies are evolving due to major mergers, and to 
measure the merger fraction at any redshift.  We discuss these issues
in \S 3.3.2.  We determine the final correction for asymmetries based on 
redshift and contamination effects, and use these
to calculate the merger fraction evolution at $z < 1.2$, which we discuss in
\S 3.3.3.  In \S 3.3.4 we use these merger fractions to parameterise
the merger history. We use these results in later subsections of
\S 3 to determine the merger rate, and the average number of mergers
galaxies of various masses undergo at different redshifts.

\subsubsection{Redshift Effects and Contamination}

The basic merger fraction ($f_{\rm m}$) is calculated as the number
of mergers selected within a given redshift bin and stellar
mass limit ($N_{\rm m}$), divided by the total number of galaxies
within the same redshift and stellar mass bin ($N_{\rm T}$),

\begin{equation}
f_{\rm m}({\rm M_{*}},z) = \frac{N_{\rm m}}{N_{T}}.
\end{equation}

\noindent However, there are several issues that have to be considered before
we can use equation (8) to derive the merger fraction.
One factor we have to account for are the effects of redshift on the
measured CAS parameters of our galaxies. These effects include
the lowering of surface brightness, and morphological
$k-$corrections, on the measurement of merger fractions using the
COSMOS and the EGS data.   We address these two corrections separately and
quantitatively.  We also discuss how to account for the contamination 
rate due to non-mergers within our sample.

Some of these issues were addressed, at least in terms of redshift effects,
by Kampczyk et al. (2007) who examined how a volume limited
sample of 1813 low redshift galaxies in the Sloan Digital Sky
Survey would appear at higher redshifts in the COSMOS field.
Kampczyk et al. (2007) simulated galaxy mergers at low redshift 
to how they would appear at high
redshift within COSMOS. They also investigated what fraction of normal galaxies
would be misidentified as mergers due to chance superpositions, and
other effects.  They found that only one in five of the merging
galaxies in the SDSS would still be identified as such if
placed at $z \sim 1$, after imaged within COSMOS. Kampczyk et al.
(2007) also found that 1.7\% of normal SDSS galaxies would be 
identified as mergers due to chance superpositions.  They however
conclude that most of the mergers they identify by eye in
COSMOS at $z \sim 1$ are real. 

We go beyond Kampczyk in this paper in a number of ways. First
we use quantitative methods that do not rely upon subjective
understanding of what a merger is.  Because we purely use the CAS
system in this paper, our approach is quantitative and repeatable, and
not subject to classification bias.  
 We also quantify directly using simulations how
much the parameters we measure are affected by redshift effects, not
just the effects of signal to noise and resolution examined by
Kampczyk et al. (2007).  We also note that Kampczyk et al. (2007)
find that the mergers they simulate remain asymmetric, but that
the Gini/M$_{20}$ method fails to locate 90\% of their merging
galaxies, similar to our findings in the Hubble Ultra Deep
Field (Paper I; Conselice et al. 2008a).

Since we use the F814W data to measure
the CAS values, and therefore merger fractions, of our galaxies, we
are sampling different rest-frame wavelengths for galaxies at different 
redshifts.  We address this empirically in
two different ways.    The first
is that we can use imaging from the Hubble Ultra Deep Field, where
we have BV$iz$ imaging for all galaxies, to determine the
morphological $k-$correction present within our galaxies, 
as discussed in Conselice et al. (2008a).  The other is
that we can use the F606W (V-band) and F814W (I-band) data
from the EGS as another determination for how the
morphological k-correction changes the resulting measured
CAS parameters.

The F814W band at observed $\lambda = 8332$ \AA\, results in a range
of probed rest-frame wavelengths from 0.7 $\mu$m at $z = 0.2$, to 0.38 
$\mu$m at $z \sim 1.2$.  This spans roughly the entire optical
wavelength range, where the CAS parameters can slight change
(e.g., Conselice et al. 2000a).  However, these resulting changes are not 
large (e.g., Taylor-Mager
et al. 2007; Conselice et al. 2008a).  As such, we investigate
the changes using galaxies in the UDF, to determine how
much the measured asymmetry in the F814W band changes due to
the rest-frame wavelength probed.

We base our measurement of the morphological $k-$correction on
how the measured CAS parameters, particularly the asymmetry
index, change with respect to $\lambda = 5500$ \AA.  For
systems at $z < 0.8$ we utilise the results of Conselice
et al. (2008a) who find a change of asymmetry parameter
with rest-frame wavelength in the optical of $\delta A_{\rm k-corr}/\lambda = 
-0.30$ $\mu$m$^{-1}$ at $z < 0.75$.  For galaxies at $z > 0.75$ Conselice
et al. (2008a) find a change of 
$\delta A_{\rm k-corr}/\lambda = -0.8$ $\mu$m$^{-1}$.  
The effective result of this
is a maximum change in the asymmetry index of $\delta A_{\rm k-corr} = 0.1$
at the highest redshifts.  We perform this correction for each of our
galaxies, such that we are measuring the CAS parameter at the same
wavelength, in this case at rest-frame $\lambda = 5500$ \AA.  We find similar
results when comparing the V and I band CAS values within the
ACS imaging of the EGS.

Another correction due to redshift is the fact that the surface
brightness declines with $(1+z)^{4}$, and thus the galaxies
we image at $z \sim 1$, with the same effective surface brightness, will
appear a factor of $\sim 8$ fainter in their measured surface brightness
than at $z \sim 0.2$.  We correct for this effect through simulations,
as described previously in Conselice (2003) and Conselice
et al. (2003a).  These simulations involve nearby galaxies which
are simulated to various redshifts, and their CAS parameters are
measured and compared with those at lower redshifts. What we find
is that up to $z \sim 1$, the CAS parameters are not affected to
a large degree, and the asymmetry index declines
by $\delta A_{\rm SB-dim} \sim 0.05$ up to $z \sim 1$.  We therefore use this
as the maximum correction at $z \sim 1$ and apply it to galaxies
at $z > 0.5$ where this effect is found to be important.

The other correction we perform is to correct the CAS merger
fraction measures for contamination from galaxies which are within
the merger region of CAS space, but which are not structurally
undergoing a merger (see also \S 3.2 \& Conselice et al. 2008a).  
The estimates of this correction must
be done by eye, and the large sample in this paper makes
this exercise too prohibitively large to carry out. Previous
studies however have revealed what this fraction is from
samples of $\sim 1000$s of galaxies (Jogee et al. 2008; Conselice
et al. 2008a). Conselice et al. (2008a) find that the
contamination rate is roughly 14$\pm$10\% for galaxies
observed in the $z-$band (Conselice et al. 2008a).  
Other results include an examination of
the CAS and morphological types for $\sim 3600$ galaxies
in the GEMS V$_{606}$ band (Jogee et al. 2008), which find a contamination rate
of $f_{\rm contam} = \sim 30$\% at $z = 0.3$ which is rest-frame 
$\sim 4500$ \AA.
Since the GEMS survey better matches the depth and rest-frame
wavelength of our images, we use a contamination correction
of $f_{\rm contam}$ = 25\%. This correction is potentially smaller than 
this due to the removal of `irregular' galaxies from Jogee et al. (2008),
some of which are potentially merging systems.  

The final
measure of the asymmetry index to calculate the merger fraction is 
given by:

\begin{equation}
A_{\rm final} = (A_{\rm obs} + \delta A_{\rm SB-dim} + \delta A_{\rm k-corr}) \times (1 - f_{\rm contam}),
\end{equation}

\noindent where $\delta A_{\rm k-corr}$ is the (usually negative) 
morphological k-correction, $\delta A_{\rm SB-dim}$ is the (positive) 
correction for redshift effects,
and $f_{\rm contam}$ is the fraction of
galaxies found through CAS which are not actually merging systems, which
we take as $f_{\rm contam}$ = 0.25 in this paper.

\subsubsection{The Merger Fraction Evolution}

The first observation we derive from our CAS values is the evolution 
of the merger fraction (Figure~5) for both the EGS and COSMOS fields
separately. Error bars reflect uncertainties due to shot noise, and
photometric redshift errors.    Before we discuss the merger fraction 
evolution it is  important to address the issue of whether our merger 
analysis, which finds galaxies with a high asymmetry, locates the same
types of galaxies at different redshifts. A legitimate concern is that
at higher redshifts, galaxies can be asymmetric due to reasons other
than major mergers, such as star formation and minor mergers.

Certainly, we know there are differences between
galaxies at various redshifts, as for example, the star formation
rate at z = 1 is several times higher than it is at z = 0.
However, several detailed analyses over the past few years have
shown that galaxies with high asymmetries are almost always major
mergers. This includes calibrating how the asymmetry index works, not 
just at  low redshifts (Conselice 2003), but also at high redshift 
(e.g., Conselice et al. 2005, 2008a).  These studies have found that galaxies 
with evidence for merging, both kinematically (e.g., Kassin
et al. 2007), and through visual inspection, have high asymmetries, 
and likewise the non-mergers are not asymmetric.
Note that this is only true in optical light, as asymmetries for
star forming galaxies in the rest-frame UV can be quite high,
independent of any merging activity (Taylor-Mager et al. 2007).

It is also possible that interactions and minor mergers can produce 
large asymmetries.  However, for reasons explained in previous papers, 
such as Conselice (2006b) and de Propris et al. (2007), it is
unlikely that very many asymmetric galaxies 
are produced through interactions or minor mergers.
de Propris et al. (2007) examined the asymmetry values for 
interacting, merging, and normal galaxies in the nearby universe.  
They found that very few of the interacting galaxies are 
considered mergers in the CAS system,
and those that would be counted are in very close
pairs which are about to merge.  By far the bulk
of the systems with high asymmetries are found 
in the major merger category.  This was furthermore
quantified through N-body simulations by Conselice
(2006) who found that only major mergers with
ratios of $> 1:4$ would be counted as a major merger
within the CAS system.  Because the asymmetry index,
and the other CAS values, are luminosity weight, faint 
features, such as tidal debris (e.g, Kawata et al. 2006)
do not produce large asymmetries.

  We thus determine the merger fraction for galaxies of 
various masses using the criteria from equation (6) and by using
the asymmetries calculated in equation (9).
Our final merger fraction 
values are tabulated in Table~1, and are separated into the EGS and
COSMOS fields in Figure~5.  Figure~6 shows the merger fraction evolution
for our M$_{*} > 10^{10}$ \solm sample, with the $z = 0.05$ point from 
De Propris et al. (2007), and the $z > 1.2$ merger fractions from 
Conselice et al. (2008a).

\vspace{1cm}
\setcounter{table}{0}
\begin{table}
 \caption{Merger Fractions for Galaxies with M$_{*} > 10^{10}$ \solm}
 \label{tab1}
 \begin{tabular}{@{}cccc}
  \hline
\hline
z & f(EGS) & F(COSMOS) & F(EGS+COSMOS) \\
\hline
0.25 & ...           & 0.05$\pm$0.01 & 0.04$\pm$0.01 \\
0.35 & ...           & 0.04$\pm$0.01 & 0.04$\pm$0.01 \\
0.45 & 0.02$\pm$0.02 & 0.04$\pm$0.01 & 0.04$\pm$0.01 \\
0.55 & 0.04$\pm$0.02 & 0.05$\pm$0.01 & 0.04$\pm$0.01 \\
0.65 & 0.09$\pm$0.03 & 0.11$\pm$0.01 & 0.09$\pm$0.01 \\
0.75 & 0.11$\pm$0.02 & 0.10$\pm$0.01 & 0.12$\pm$0.01 \\
0.85 & 0.08$\pm$0.02 & 0.12$\pm$0.01 & 0.11$\pm$0.01 \\
0.95 & 0.09$\pm$0.02 & 0.10$\pm$0.01 & 0.10$\pm$0.01 \\
1.05 & 0.15$\pm$0.02 & 0.11$\pm$0.01 & 0.11$\pm$0.01 \\
1.15 & 0.15$\pm$0.03 & 0.12$\pm$0.01 & 0.13$\pm$0.01 \\
\hline
\end{tabular}
\end{table}

We note that the merger fractions we measure are in no sense the total 
galaxy merger 
fraction, that is the total number of galaxies which have undergone, or are
undergoing, a merger at the given time.    Most galaxies will have undergone
a merger sometimes in their history, and the fraction of galaxies which
have undergone a merger sometime in the past will be close to 100\%.
 
All merger fractions we derive in this paper are computed using the 
CAS parameters,
which is only sensitive to a well defined time-range during the
major merger process (Conselice 2006; Lotz et al. 2008b). In the case of 
CAS mergers, this time-span is roughly 0.4-1 Gyr (Conselice 2006; Lotz
et al. 2008b). As shown in Conselice (2006) there are phases of a merger 
which will not be picked up by the CAS technique.  A different technique 
will find a different merger fraction if it has a time sensitivity different
from the CAS system. For example if galaxy pair methods has
roughly a factor of two longer time-scale for finding a merger
than the CAS system, the resulting computed merger rate would be the
same, as the pair method would find a factor of two more galaxies merging
than the CAS parameters. We therefore expect, and find there to be, 
galaxies that by eye 
appear as a merger, but do not have a high asymmetry (e.g., Conselice
et al. 2008a).  

There are a few obvious features of the merger fraction that
deserve note, as seen in Figure~5 and 6.   The first is that
at the redshifts where we have data for both the EGS and COSMOS,
the two agree remarkably well, always within 1-$\sigma$ (see also
Table~1). There is also
a general trend for the merger fractions, as measured in both fields,
to decline at lower redshifts, going from $f_{\rm m} = 0.15\pm0.03$
at $z = 1.2$ down to $f_{\rm m} = 0.02\pm0.02$ at $z = 0.4$ in the EGS 
field, with a similar trend within the COSMOS field.  
The lowest
redshift point for the COSMOS data at $z = 0.2$ is slightly higher than
the two nearest and higher redshift bins. This is at least partially
due to the higher resolution reached at this redshift, which can
result in higher computed asymmetries (e.g., Conselice et al. 2000a).

The other obvious and outstanding feature of our  measured
merger history is the fact that the merger fraction, based on our
measurements, declines rapidly at $z < 0.7$, although it
appears roughly constant at $0.7 < z < 1.2$.  This result is
unlikely solely due to cosmic variance, as both fields show this
behaviour, and this result is relatively insensitive to the
exact corrections we apply for redshifts and $k-$corrections.
This result is, within our errors, significant at 
$> 4$-$\sigma$, based on the combined EGS+COSMOS sample.  

We also
fit these two redshift ranges in the EGS+COSMOS merger fraction
evolution separately, as shown in Figure~6
by the two dashed lines. Here we fit both the evolution at
$z < 0.7$ and $0.7 < z < 1.2$ as power-laws, in which we find
very different slopes. At $z < 0.7$ we find that the best fit
power-law slope is $m = 5.2\pm1.0$ with $f_{0} = 0.009\pm0.003$,
going right through the de Propris et al. (2007) $z = 0.05$
value.  At $0.7 < z < 1.2$ the slope is $m = 0.07\pm0.55$,
which indicates very little evolution, and therefore must
be the result of significant merging activity within this
epoch.  We also see a turnover in the merger history at
$z > 2$ (Conselice et al. 2008a), which has also been seen
for galaxies in pairs at similar redshifts (Ryan et al. 2008).

Furthermore, Figure~7 compares our merger fractions with results from
other merger fraction studies in Figure~7.
Specifically, we compare our merger history with the results 
of both previous pair and structural studies.  Points on
Figure~7 include de Ravel et al. (2008) who
use the VVDS to determine the pair fraction evolution for
galaxies from $z = 0.5$ to $z = 1$.  We plot on Figure~7
the results from this study for galaxies which have
a pair separation of $<$ 20 h$^{-1}$ kpc, and a velocity
difference $\delta V <$ 500 \kms\, within a magnitude range
of M$_{\rm B} < -18 - Q(z)$.  The $Q(z)$ factor accounts for the
evolution of stellar populations, and is an attempt to match
the photometric evolution of galaxies so as to obtain the
same galaxies at different redshifts. We also compare with recent
DEEP2 results for the pair fraction evolution from Lin
et al. (2008), who find similar pair fractions to
de Ravel et al. (2008).  The other pair fraction we compare
with is Kartaltepe et al. (2007) who measure pair fractions
within COSMOS within $<$ 20 h$^{-1}$ kpc separation using photometric
redshifts.  This paper finds 1749 galaxy pairs for
systems brighter than M$_{\rm V} = -19.8$, roughly equivalent
to the stellar mass limit we have used.  Finally, we compare
our merger fractions to those published by Lotz et al. (2008a)
using the Gini/M$_{20}$ method for 3009 galaxies within the EGS 
brighter than 0.4L$^{*}_{\rm B}$.

\begin{figure}
 \vbox to 120mm{
\includegraphics[angle=0, width=90mm]{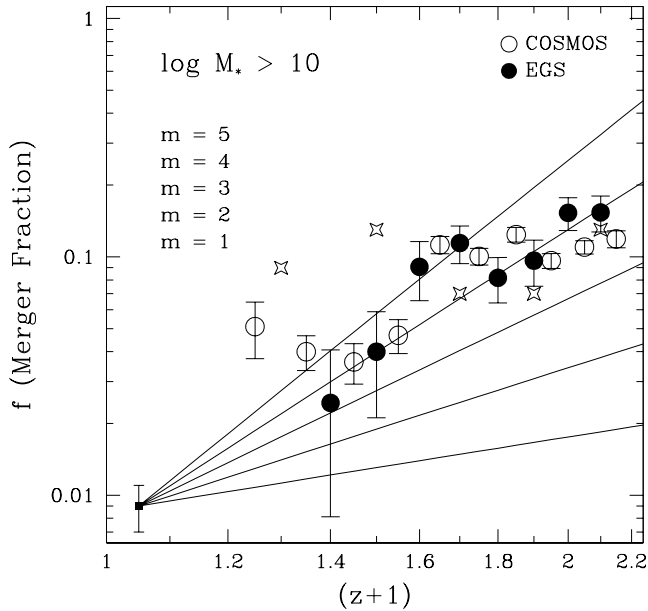}
 \caption{The evolution of the derived merger
fraction up to $z = 1.2$ through the use of the CAS system for
the EGS (solid circles) and COSMOS (open circles)
fields.  The open stars are the derived merger
fractions from the EGS using the Gini/M$_{20}$
parameters from Lotz et al. (2008).  The point
at z = 0.05 originates from De Propris et al. (2007).  The five
solid lines show the evolution from $z = 0.05$ as a power-law,
with $f_{0}$ fixed to 0.009, and with various values of the power-law
index shown; m = 1,2,3,4,5. }
\vspace{3cm}
} \label{sample-figure}
\end{figure}

\begin{figure}
 \vbox to 140mm{
\includegraphics[angle=0, width=90mm]{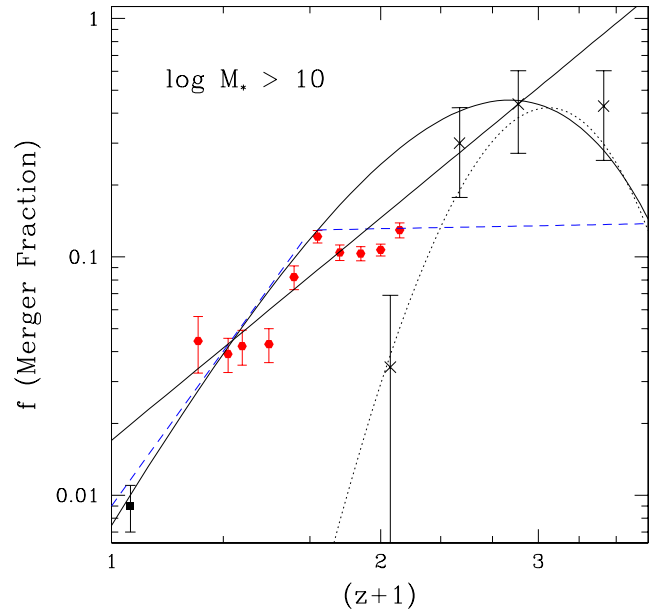}
 \caption{The evolution of the merger fraction from
$z = 0$ to $z = 3$ using structural parameters from
the CAS system.  The point at $z = 0.05$ originates
from the study of De Propris et al. (2007).  The points
between $z = 0.4$ and $z = 1.2$ are from the combined EGS
and COSMOS fields, which are plotted individually
on Figure~5.  The points from $z > 1$ (crosses)
are taken from the CAS study of the Hubble Deep and Ultra
Deep Fields (Conselice et al. 2008a).  We show several
fits as well, with the solid straight line the best fit 
power-law, while the curved solid line is the best fit
for the combined exponential/power-law parameterisation
(see text).    Also shown as the two dashed blue lines are
the best fit power-law parametrisation for the EGS+COSMOS
points at $0.2 < z < 0.7$ and $0.7 < z < 1.2$, fit separately.
This shows the rapid evolution in the merger fraction within
$z < 1.2$.   The dotted line shows the previous
best fit exponential/power-law relation for galaxies
with stellar masses M$_{*} > 10^{10}$ \solm described
in Conselice et al. (2008a).  }
} \label{sample-figure}
\vspace{1.5cm}
\end{figure}

\begin{figure}
 \vbox to 140mm{
\includegraphics[angle=0, width=90mm]{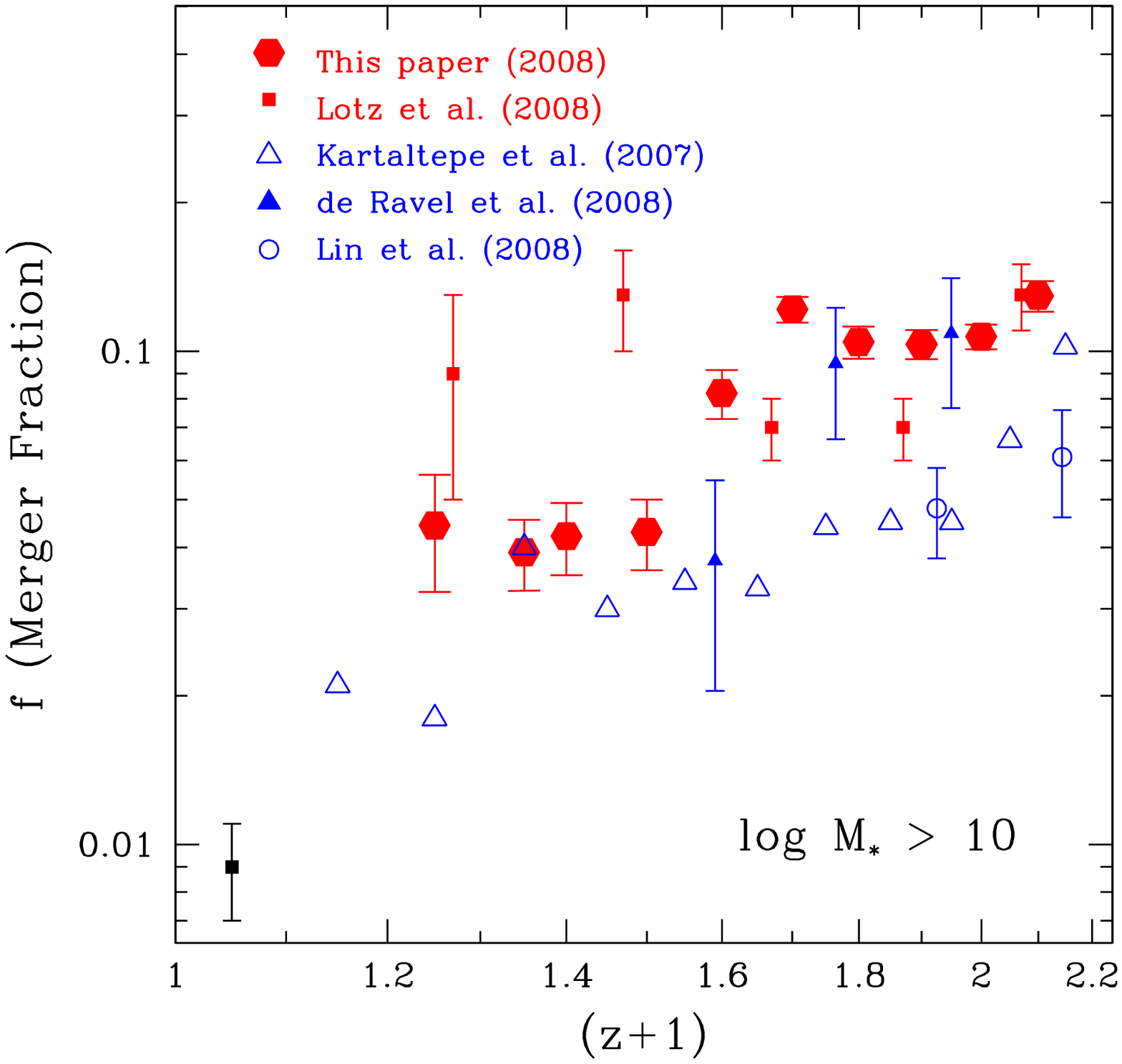}
 \caption{The evolution of the derived merger fraction
through several previous studies compared to our results.
Those systems which are coloured red are those derived
through structural methods, either the CAS system (this
paper) or the Gini/M$_{20}$ method (Lotz et al. 2008a).  
The blue symbols are the merger fractions derived from
pair studies, either kinematic pairs as in Lin et al.
(2008) and de Ravel et al. (2008) or photometric pairs
from the COSMOS fields (Kartaltepe et al. 2007).  In
general the morphologically defined mergers give a higher
merger fraction than those derived through pairs. }
} \label{sample-figure}
\end{figure}

\subsubsection{Extremely Massive Galaxies}

In addition to systems selected by M$_{*} > 10^{10}$ \solm,
we also examine the evolution of the merger fraction for
the extremely massive galaxies in our sample, those
with M$_{*} > 10^{11}$ \solm.  There are several issues
however when measuring the merger fraction for these
systems. The first is simply that there are not nearly
as many galaxies with these masses as for systems selected
by M$_{*} > 10^{10}$ \solm.  This can be seen in Figure~8,
where at the lowest redshifts at $z \sim 0.3$ there are not
enough systems to measure the merger fraction, and very few at
$z \sim 0.4$, with resulting large error bars.

In any case, what we find is that the merger fractions for
M$_{*} > 10^{11}$ \solm galaxies at $0.2 < z < 1.2$ overall
are very similar to that for M$_{*} > 10^{10}$ \solm systems.
It appears that at $z > 0.7$ the merger fraction for these
extremely massive galaxies is slightly lower than for the
$10^{10}$ \solm systems.  This changes at lower redshifts
where the M$_{*} > 10^{11}$ \solm galaxies appear to have slightly
higher merger fractions, although we note that the error bars on these 
fractions
do not rule out that these points are in fact as low, or even
lower than, those for systems at $10^{10}$ \solm.

\subsubsection{Parameterisation of the Merger Fraction Evolution}

There are a few popular ways to fit the merger fraction evolution.
The first is the traditional power-law format (Patton et al. 2002; Conselice
et al. 2003; Bridge et al. 2007) which dates back to early work
by Zepf \& Koo (1989) on the evolution of the merger fraction.  
This fitting format is given by,

\begin{equation}
f_{\rm m}(z) = f_{0} \times (1+z)^{m}
\end{equation}

\noindent where $f_{\rm m}(z)$ is the merger fraction at a given
redshift, $f_{0}$ is the merger fraction at $z = 0$, and
$m$ is the power-law index for characterising the merger
fraction evolution.  Zepf \& Koo (1989) calculated a merger
fraction evolution which increased with redshift up to $z = 0.25$, and
obtained a slope of $m = 4.0\pm2.5$. Nearly all further
studies have found values within this range. 

However there
is currently a debate over the true nature of the increase, with
some studies finding a fast evolution (Conselice et al. 2003; Kartaltepe et
al. 2007), while others have found a more modest evolution
(e.g., Lotz et al. 2008a; Lin et al. 2008).  There are several
reasons for this diversity in the parameterisation of the merger fraction
evolution, which we discuss after demonstrating the various ways the
merger fraction evolution can be fit within our own sample.

We explore below the power-law behaviour of the merger fraction evolution
in a few  ways. The first method is to investigate the
power-law fit to the merger fractions solely within the COSMOS+EGS
data set (Figure~5).   When we fit a power-law to this evolution using
just the points from COSMOS+EGS at $0.2 < z < 1.2$
we find a best fit given by $f_{0} = 0.025\pm0.005$ and $m = 2.3\pm0.4$.  That
is, internal to itself, the CAS merger fraction from $z = 0.2$
to $z = 1.2$ increases modestly with time.   Note that the
derived merger fraction at $z = 0$ for this sample is 0.025$\pm0.005$
which is about a factor of three larger than the $z = 0$ merger
fraction derived from de Propris et al. (2007).  When we hold the $z \sim 0$
point at a constant value of $f_{0} = 0.009$ we find that the best-fit
slope for COSMOS+EGS data increases to $m = 3.8\pm0.2$.

Our derived evolution is similar to the results found by Lotz et al. (2008a), whose merger fractions
are similar to ours (Figure~7). Our merger fraction evolution power-law slope,
$m$, is however smaller than the index $m$ measured
by Kartaltepe et al. (2007) using pair counts.  Even
though Kartaltepe et al. (2007) find in general lower pair
fractions compare to our merger fractions, the increase with time
is more pronounced, and they derive an index of $m =  3.1\pm0.1$. This is more
than 2-$\sigma$ away from our result, and this difference deserves some
explanation.  

When we fit the  Kartaltepe et al. (2007) points
with the same $z = 0$ prior we use for the EGS+COSMOS data, we find
that the best fit slope is $m = 2.9\pm$0.1.  When we remove the
$z \sim 0$ prior in this fit, we find that the slope changes
to $m = 2.7\pm0.6$, although most of this is driven by the high
redshift points from Kartaltepe et al. (2007).  If we remove the
highest redshift point (Figure~7), the measured value of $m = 1.8\pm0.3$, which
is very similar to our own value at the same redshift range.


\begin{figure}
 \vbox to 140mm{
\includegraphics[angle=0, width=90mm]{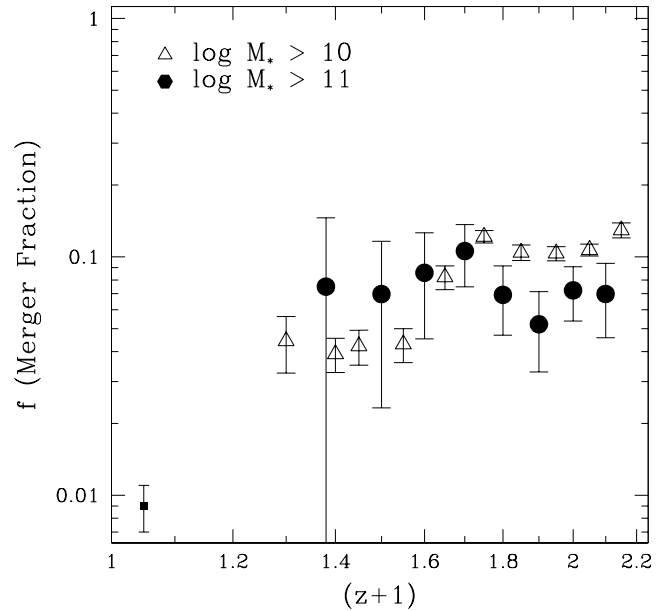}
 \caption{A comparison between the evolution of the merger
fraction using the CAS method for galaxies with M$_{*} > 10^{10}$ \solm
and M$_{*} > 10^{11}$ \solm from $z = 0.2 - 1.2$.  In general we
find that the evolution at higher masses is similar to those presented
in the other plots in this paper, although the higher mass systems
have a higher uncertainty associated with their measured merger
fractions. }
} \label{sample-figure}
\end{figure}

Overall, we find that the value of $m$ can vary significantly depending
on what value of $f_{0}$ is chosen. For example, if the $z = 0$ 
merger fraction
for log M$_{*} > 10$ galaxies is twice as high as the de Propris 
et al. value (i.e., $f_{0} = 0.018$) then the best fit value of
the power-law slope changes from $m = 3.8\pm0.2$ to $m = 2.8\pm0.1$, thus
raising the $z = 0$ merger fraction lowers the value of the power-law
slope.   If we hold the $z \sim 0$ merger fraction as 4\%, then we find
that the power-law slope falls to $m = 1.5\pm0.1$.  Note that
this does not appear to be the case for very massive galaxies
with M$_{*} > 10^{11}$ \solm (Bluck et al. 2008).

Since the $z \sim 0$ merger fraction can be critically important for
deriving the merger history if it is used as a prior, it is worth 
examining what the merger fraction
at $z \sim 0$ is, and how certain we know this number. We also have some 
estimates of what the local merger fraction is from previous
work such as Patton et al. (2000) and de Propris et al. (2007). We use the 
Millennium Galaxy Catalog (MGC) selection for mergers, which is based
on systems which are at $M_{\rm B} < -18$, and thus a complete comparison
is not possible without measuring the stellar masses for these
galaxies.   De Propris et al. (2007) find that the merger fraction at 
$z \sim 0.05$, is $f_{0} = 0.009\pm0.02$, based on the CAS asymmetry.  
By holding the value of $f_{0}$ fixed
to 0.009, we fit a power-law slope of $m = 3.8\pm0.2$.  However,
this is not a good fit to the data (Figure~6), and tends to under-predict
the merger fraction at $z = 0.8$ and over-predict the fraction at
$z > 0.8$.  

One of the features of the derived merger fraction up to $z = 1.2$
is that between various neighboring redshifts there are
significant differences in the derived merger fraction (Figures~5-7).
This is true for the combined EGS+COSMOS sample, the EGS and COSMOS
samples alone, and is also found for the Gini/M$_{20}$ approach
in Lotz et al. (2008a), as well as the pair method derived
in Kartaltepe et al. (2007).  In fact, the slope of the power-law
merger fraction evolution ranges from $m = 5$ to $m = -3$ between
two neighboring individual merger fractions.
It is not yet clear what the origin of these differences are, whether the
result of cosmic variance, or systematics. However on average the
merger fraction does increase slightly with time up to $z = 1.2$
(\S 3.3.3).

We do a similar examination of the best power-law fit to the
merger fraction evolution by including  higher redshift merger
fractions taken from a combined UDF+HDF sample (Conselice et
al. 2008a) selected with the same stellar mass cut of
M$_{*} > 10^{10}$ \solm.  The resulting merger fraction evolution
is shown in Figure~6, which is currently our best estimate
of how the merger fraction varies with time for a mass selected
sample.  When we fit the merger fraction evolution for 
M$_{*} > 10^{10}$ \solm galaxies from $z = 0$ to $z = 3$ we find that the 
best fit using all points, and no priors, is: $f_{0} = 0.015$ and 
$m \sim 3$, although this power-law fit is poor, and the $\chi^{2}$ is
large.

If we hold the value of $f_{0} = 0.009$, we find that the best
fit value of the power-law slope increases to $m = 3.7$, although this
does not go through the highest redshift point, and results in a
poor fit.   However, the $z = 0$ extrapolation from the best
fit to all the data predicts that the nearby merger fraction
is roughly twice the best value we currently have.   This again
demonstrates the importance of the value of $f_{0}$ within
the power-law fit.  The value of $f_{0}$ determines to a large
degree what the fitted slope of the power-law evolution will be,
especially if the value is held constant during the fit.


Another way to characterise the merger fraction evolution, which
dates back to theoretical arguments based on the Press-Schechter formalism
for merging (Carlberg
1990), is a combined power-law exponential evolution.  This form
appears to be a better fit to all of the redshift data than a simple
power-law (Conselice 2006b).  This is however not the
case for the merger fraction for M$_{*} > 10^{11}$ \solm galaxies at $z < 3$, 
which can be fit by a power-law (Bluck et al. 2009).  The formula for the
power-law/exponential evolution is given by:

\begin{equation}
f_{\rm m} = \alpha (1+z)^{m} \times {\rm exp}(\beta(1+z)^{2}),
\end{equation}

\noindent where the $z = 0$ merger fraction is given by
$f_{\rm m}(0) = \alpha \times$ exp($\beta$).  We find in
general that this combined exponential power-law fit is
better than a simple power-law and is likely a better
representation for how the merger fraction evolves with time.
We also find this to be the case for a power-law/exponential
form, without the square on the $(1+z)$ term in the exponential.
However, none of these forms are very satisfying, given
the large errors on their fits, and the three free 
parameters needed to create the fit.

As discussed in Paper II (Conselice et al. 2009, submitted) 
there does not appear to be a simple way 
to parameterise the merger fraction.  We explored several fitting
routines, and found that the best-fit two parameter model
is an exponential/power-law of the form:

\begin{equation}
f_{\rm m} = \alpha (1+z)^{3} \frac{1}{{\rm exp}(\beta \times z)},
\end{equation}

\noindent which is designed such that $f_0 = \alpha$, and 
is only a two parameter
parameterisation, and fits the data as well as the three parameter
model above.  In any case, we have found that the fitting
with the exponential/power-law and the Carlberg (1990) version of
the exponential/power-law, gives an exponent on the power-law
portion of $m \sim 3$.  This is the index predicted for how the
number densities of galaxies decline with redshift, and is
therefore the natural exponent on the power-law for a passive
merger evolution, which appears to occur at $z < 0.7$.




\subsection{Comparison of Structural Mergers and Pair Fractions}

One of the results we derive from our measured structural
mergers is how the derived merger fractions, based on the CAS and
other methods (such as Gini/M$_{20}$), compare with derived pair
fractions.    Figure~7 shows this comparison, with the morphological
methods (this work \& Lotz et al. 2008a) shown in red, and the
pair methods (Kartaltepe et al. 2007; Lin et al. 2008; de Ravel
et al. 2008) shown in blue.  What is immediately clear is
that the structural methods, while agreeing quite well with
each other, find a higher merger fraction than the pair
methodology.  There are several possible reasons for this.

While it is possible that one or both methods have systematics
that result in inaccurately measured merger fractions, we first 
investigate what the
time-scales for these methods are.  This is an important
question as these fractions differing can be partially, or entirely,
explained
if the time-scales for merging for the two methodologies are
different. For example, if the time-scale for a 20 kpc pair
to merge is half the time-scale sensitivity for an asymmetric
galaxy, then we would expect the pair fraction to be half of
the CAS merger fraction if both methods are tracing the same 
merger process.  That is, the merger fraction for a given
sample scales as the time-scale sensitivity of the 
method used to find mergers.  Since the merger rates for the
two methods should give the same result, then the ratio of
the pair fraction to its merger timescale should be equal
to the structural merger fraction divided by its time-scale,
or:

\begin{equation}
\frac{f_{\rm pair}}{\tau_{\rm pair}} = \frac{f_{\rm CAS}}{\tau_{\rm CAS}}.
\end{equation}

\noindent Time-scales ($\tau$) within the merger process are notoriously 
difficult
to measure, and have large uncertainties.  Initially, measuring
the merger rate from pairs involved dynamical friction calculations
(e.g., Patton et al. 2002; Conselice 2006), and typical time-scales
for a 20 kpc pair to merge are 0.5-1 Gyr with various assumptions.
The time-scale calculation for dynamical friction used by Conselice (2006b), 
and in earlier studies are based on isothermal mass distributions and the
time-scale can depend highly on the mass of the galaxies, and the 
characteristic velocity of the system (e.g., Conselice 2006b, eq. 7).

Likewise, the time-scales for merging within the CAS system for
dark matter dominated galaxies was found to be similar to the 
dynamical friction time-scales derived from an isothermal profile,
and a galaxy with mass of $\sim 10^{10}$ \solm.  There are however
problems with both of these calculations, which have already
been alluded to above for the dynamical friction time-scale.
The measured time-scales for the CAS method are found by
Conselice (2006) to vary between $\sim 0.3 - 0.8$ Gyr, depending
on viewing angle and the orbital parameters of the two galaxies
in the pair.  Also, the simulations used in Conselice (2006) are
purely dark matter, and it is desirable to determine the
CAS time-scale for systems with stars, star formation, and dust.

New simulations were recently analysed by Lotz et al. (2008b) in
terms of CAS, Gini/M$_{20}$, and pair selection for mergers.  Lotz et al. (2008b) furthermore
utilise simulations that include star formation and dust, and are
currently by far the most thorough investigation into merger
time-scales using both morphology and the time-scales for mergers
to occur within a given pair separation from 20 h$^{-1}$ kpc to 
30 h$^{-1}$ kpc, and 50 h$^{-1}$ kpc.  

Lotz et al. (2008b) find a variety of time-scales for their merger
simulations depending on the type of merger and the type of galaxy.
For their highest resolution simulation (SbcPPx10), they calculate
a merger time-scale of $\tau_{\rm CAS} = 0.94 \pm 0.13$ for the
CAS methodology, and a time-scale of $\tau_{\rm pair} = 0.15 \pm 0.18$
for galaxies to merge within a 20 h$^{-1}$ kpc pair.  This ratio of
time scale, which we denote as $\kappa = \tau_{\rm CAS}/\tau_{\rm pair}$,
is equal to $\kappa =$ 6.3 for this model.  

Figure~9 shows the histogram
for the value of $\kappa$ for all the models published in Lotz et al.
(2008b).  What we find is a general distribution, but with all the
Sbc models having a ratio $\kappa > 1$.  The models shown by the
blue hatched histogram in Figure~8 are for the Lotz et al. `G' models,
which are less gas dominated, and have sub-parabolic orbits which
lead to artifically shortened merger-times in pairs.  As can be
seen, while earlier analytical calculations showed that the time-scale
for pair merging was similar to the morphological merger time-scale,
the simulations by Lotz et al. demonstrate that the time-scales
for pairs to merge are shorter than the analytical estimates.

This implies that if the pair method and the structural methods 
are measuring the same merger process, then the merger fractions
derived from pairs should be lower than those derived from structure
by an equivalent amount.  We denote this ratio as 
$\kappa' = f_{\rm CAS}/f_{\rm pair}$.  For the COSMOS field
we find that the value of $\kappa'$ using the data from
Kartaltepe et al. (2007) and this paper, give $\kappa'$ values
from 1.5-3.5, yet the pairs from Kartaltepe
et al. (2007) are not selected in the same way our CAS mergers
are, i.e., with M$_{*} > 10^{10}$ \solm.

A better test of the merger criteria time-scale is to determine how
the ratio of the CAS merger fraction and the pair fraction change
with redshift in the same sample and stellar mass selection.  
While it is likely that
even within the same sample, the CAS and pair methods will find
galaxies in different modes of evolution (De Propris et al. 2007), it
is still instructive to determine this ratio within a well defined
sample. We utilize the POWIR database from the EGS to determine
this ratio (e.g., Conselice et al. 2007b, 2008b).  

Figure~10 shows this ratio for galaxies at separations of less
than 20 kpc, 30 kpc, 
and 50 kpc.  The values for f$_{\rm m}$(CAS)/f$_{\rm m}$(Pair),
with a pair defined as having a separation of $< 20$ kpc, range
from 12.2 to 2.3, with a average value of 6.2, which perhaps coincidently,
is close to the value of $\kappa$ for the highest resolution model from
Lotz et al. (2008b).  This implies that there is no inconsistency
between the pair method and structural methods for measuring 
merger fractions and rates.

\begin{figure}
 \vbox to 140mm{
\includegraphics[angle=0, width=90mm]{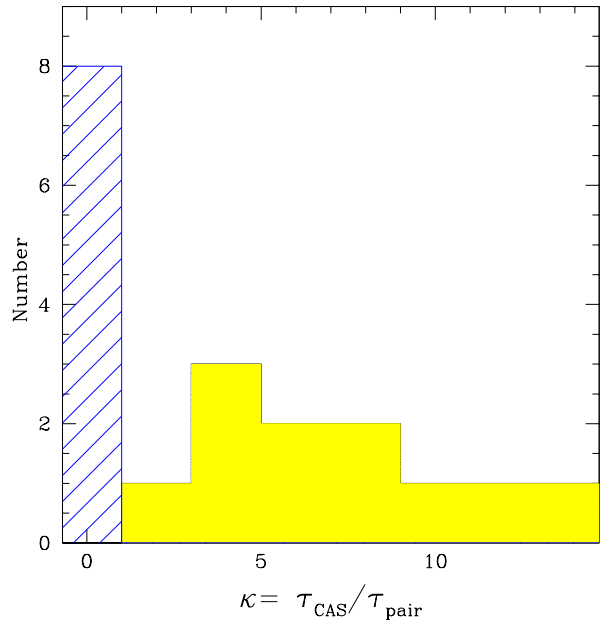}
 \caption{The ratio of the time-scale sensitivity for
the CAS identified mergers and the time-scale for merging
for galaxies in 20 h$^{-1}$ kpc pairs ($\kappa$).  The blue hatch
histogram shows the results for galaxies in sub-parabolic
orbits, while the solid yellow show the results for the
more realistic Sbc orbital models from Lotz et al. (2008b). }
} \label{sample-figure}
\end{figure}

\begin{figure}
 \vbox to 140mm{
\includegraphics[angle=0, width=90mm]{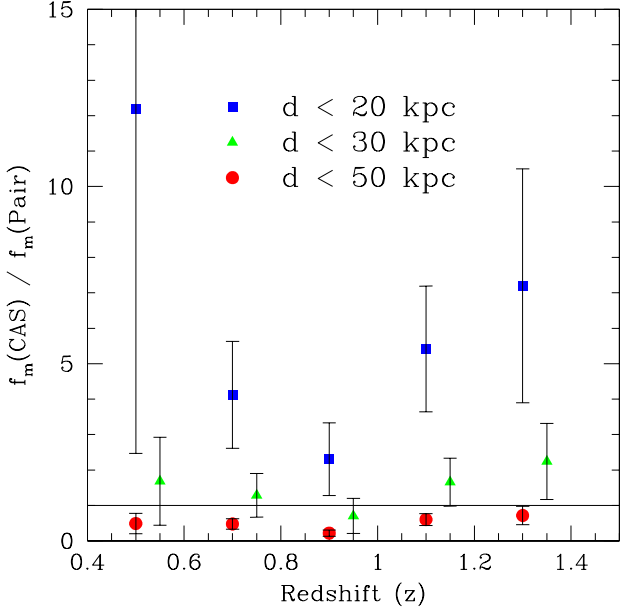}
 \caption{The ratio of the CAS merger fraction and the pair
fraction, within the same sample of galaxies with 
M$_{*} > 10^{11}$ \solm within the EGS and the other
POWIR fields (Conselice et al. 2007b, 2008b).   The blue squares
are for those systems with separations of $< 20$ kpc, the
green triangles are for those with separations of $< 30$ kpc,
and the red circles are for systems with separations of $< 50$ kpc.  }
} \label{sample-figure}
\end{figure}

\subsection{Merger Rates}

We are now in a position to measure the galaxy merger rate
for M$_{*} > 10^{10}$ \solm galaxies from $z = 0$ to $z = 3$.
The merger rate is difficult to measure, however, and our
attempt should be viewed as a preliminary full measure of
this evolution up to $z = 3$.  Measurements of the merger
rate will improve as our understanding and knowledge of
galaxy number densities, merger fractions, and merger
times scales improve. 

\subsubsection{Merger Rates per Galaxy}

 Since the merger rate has a 
relatively high error associated with it, we first
investigate the merger rate per galaxy, which is
simply just the number of mergers a galaxy of a given
mass will undergo as a function of time.  
We, in fact,
examine the inverse of this, which we call $\Gamma$,
which is the average amount of time a galaxy exists before
undergoing a merger as a function of redshift, or

\begin{equation}
\Gamma = \frac{\tau_{\rm m}}{f_{\rm gm}},
\end{equation}

\noindent where we have utilised the galaxy merger fraction,
or rather the fraction of galaxies merging in a population,
which is related to the merger fraction ($f_{\rm m}$) by:

\begin{equation}
f_{\rm gm} = \frac{2 \times f_{m}}{1+f_{m}}.
\end{equation}

\noindent In equation (14), we also use the time-scales for mergers
($\tau_{\rm m}$), which is potentially the largest uncertainty
when trying to derive evolution from the merger fraction (\S 3.4).  The value
for $\tau_{\rm m}$ is measured to range from $\tau = 0.4 - 1$ Gyr
based on N-body models from Conselice (2006b) and from Lotz et al. (2008b).
The nominal value we use, based on the average CAS time-scale using
the Lotz et al. (2008b) SbcPP, SbcPR and SbcRR simulations is 1 Gyr, with
an uncertainty in this value of 0.3 Gyr.  Note that this is
more than a factor of two larger than the time-scale used in
Conselice et al. (2008a), which utilised simulations from
Conselice (2006) to derive a merger time-scale of
0.34 Gyr.  The time-scale for a pair to merge, as found
by Lotz et al. (2008b), is 0.2$\pm0.1$ Gyr for a 20 kpc
pair, and $\tau_{\rm m} = 0.4\pm0.2$ Gyr for merging 
systems within the Gini-M$_{20}$ system, which we also
use to calculate the value of $\Gamma$, as plotted
in Figure~11.

As can be seen in Figure~11, the typical value for $\Gamma$ is around 
$\sim 10$ Gyr at $z < 1$, demonstrating that $z < 1$ galaxies
with masses M$_{*} > 10^{10}$ \solm only undergo,
on average, a single merger during this time.  We
can calculate the exact number of mergers that occur at $z < 3$ by 
integrating the inverse of $\Gamma$ over redshift,

\begin{equation}
N_{\rm merg} = \int^{t_2}_{t_1} \Gamma^{-1} dt = \int^{z_2}_{z_1} \Gamma^{-1} \frac{t_{H}}{(1+z)} \frac{dz}{E(z)},
\end{equation}

\noindent where $t_{H}$ is the Hubble time, and $E(z) = [\Omega_{\rm M}(1+z)^{3} + \Omega_{k}(1+z)^{2} + \Omega_{\lambda}]^{-1/2}$ = $H(z)^{-1}$. 
In this case, we parameterise $\Gamma$ by:

\begin{equation}
\Gamma(z) = \Gamma_{0} (1+z)^{m},
\end{equation}

\noindent where we find a best fit of $\Gamma_{0} = (13.8\pm3.1$) Gyr,
and $m = -1.6\pm0.6$.
Using equation (16), and the parameterisation of $\Gamma$ in equation
(17) we calculate the number of mergers a galaxy with 
M$_{*} > 10^{10}$ \solm undergoes from $z = 3$ to $z = 0$.
The total number of mergers  
depends strongly on the adopted value of the CAS merger time-scale
($\tau_{\rm m}$).  The range in the total number of mergers a galaxy
undergoes at $z < 3$ ranges from $N_{\rm merg}$
= 2.3 to 6.6, depending on the time-scale used.   By integrating the
individual merger fractions, we calculate that the total number of mergers
a galaxy undergoes can be expressed as $N_{\rm merg} = 2.3 \tau_{m}^{-1}$
at $z < 3$.
Most of the merging within these massive galaxies occurs at $z > 1$, 
independent of the value of the merger time-scale, as discussed in Conselice 
et al. (2008a).  

The total number of cumulative mergers a galaxy undergoes from $z = 3$
to $z = 0$ is shown in Figure~12 for three different time-scales of
CAS sensitivity to the merger process - $\tau_{\rm m} = 0.35, 0.5, 1$ Gyr.
Also shown in Figure~12, as the dashed line, is the evolution in the total
cumulative number of mergers for a population with a constant 
$\Gamma = 1$ Gyr.    Although the
total number of mergers depends strongly on the still uncertain time-scale
for the merger process, it is clear that most merging for massive
galaxies must occur at $z > 3$.  The horizontal line's intersection with
the various models shows at which redshift a galaxy will on average
have had a single major merger since $z = 3$.  This varies between $z = 1.5$ to
$z = 2.5$.  Thus, independent of the time-scale used, on average,  
massive galaxies with log M$_{*} > 10$ undergo a merger 
between $z = 2.5$ and $z = 1.5$.  Likewise, at
$z < 1$ (the vertical solid line), a typical massive galaxy will undergo 
between 0.5 to 2 major mergers, depending on the time-scale sensitivity. Our
best estimate for the amount of stellar mass galaxies with M$_{*} > 10^{10}$
\solm grow by at $z = 1$ to $z = 0$, based on the evolution of the
mass function, is a factor of $\sim 2$ increase (Conselice 
et al. 2007), demonstrating consistency between our findings and
the actual evolution in stellar mass in galaxies.

\subsubsection{The Galaxy Merger Rate}

Another important physical quantity to obtain when studying
galaxy mergers, is the galaxy merger rate ($\Re_{\rm g}(z)$), which is
the number of galaxies merging, per unit time, per
unit co-moving volume. The galaxy merger rates for our sample are calculated 
through the merger rate equation,

\begin{equation}
\Re_{\rm g}(z) = {\rm f_{\rm gm}}(z) \cdot \tau_{\rm m}^{-1} {\rm n_{\rm gm}}(z)
\end{equation}

\noindent where n$_{\rm gm}$ is the number density of
galaxies within a given stellar mass range, and f$_{\rm gm}$ (eq. 15) is the 
galaxy merger fraction. Note that
this is not the merger fraction, which is
the number of mergers divided by the number of galaxies, which is roughly 
half the galaxy merger fraction (Conselice 2006).   We obtain our galaxy
number densities $n_{\rm gm}$ from Conselice et al. (2007b) for galaxies
with stellar masses M$_{*} > 10^{10}$ \solm at $z > 0.2$, and from 
Cole et al. (2001) for galaxies at $z \sim 0$.

\begin{figure}
 \vbox to 140mm{
\includegraphics[angle=0, width=90mm]{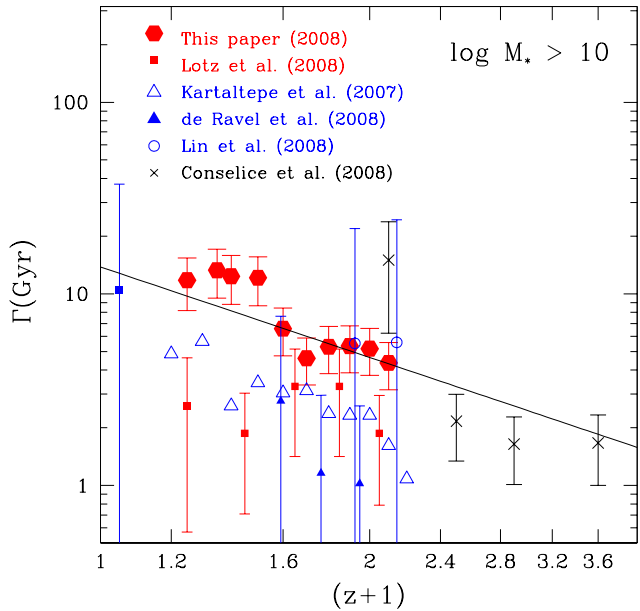}
 \caption{The evolution of the quantity $\Gamma$, which is the
average time between a merger for galaxies with M$_{*} > 10^{10}$ \solm
at $z < 3$.  We include on this plot the measured pair fractions
(e.g., Kartaltepe et al. 2007; de Ravel et al. 2008 and
Lin et al. 2008).  The merger fractions by which the value
of $\Gamma$ is calculated from galaxy structure, include
those from this paper, and previous work by Lotz et al. (2008a) at $z < 1.2$,
and Conselice et al. (2008a) for systems at $z > 1$.  The solid line
shows the best fit power-law evolution, which we use to parameterise
$\Gamma$ for measuring the cumulative merger history for 
M$_{*} > 10^{10}$ \solm galaxies.}
} \label{sample-figure}
\end{figure}

\begin{figure}
 \vbox to 140mm{
\includegraphics[angle=0, width=90mm]{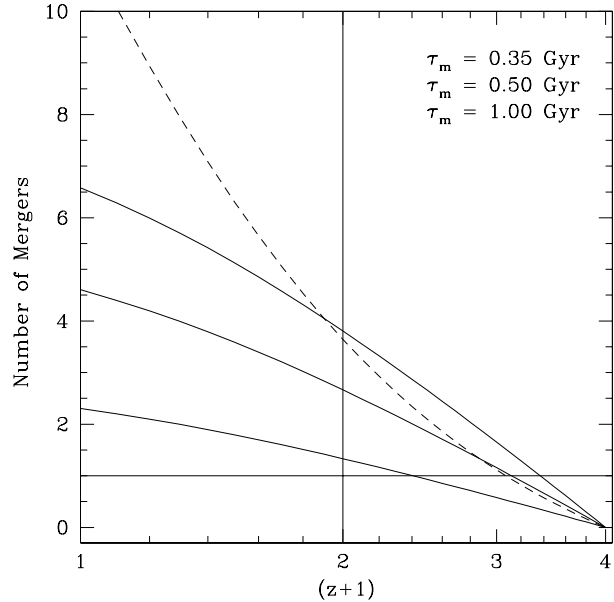}
 \caption{The integrated number of mergers since $z = 3$,
as a function of the time-scale for CAS sensitivity to the
merger process.  The three solid
lines show the evolution of how many mergers have occurred for
galaxies with M$_{*} > 10^{10}$ \solm since $z = 3$ using different
values for the time-scale in which the CAS system is sensitive
to mergers (see text).  The dashed line shows the evolution for
mergers with a constant time-scale of $\Gamma = 1$ Gyr. }
} \label{sample-figure}
\end{figure}

The resulting galaxy merging rate is shown in Figure~13.  The galaxy 
merger rate increases with time for the
M$_{*} > 10^{10}$ \solm galaxies, from $z = 0$ to $z = 3$.  This slightly
differs from the findings of Bluck et al. (2009) who find that the
merger rate for  M$_{*} > 10^{11}$ \solm galaxies is consistent with
being constant from $z = 0$ to 3.  We try to
parameterise this evolution of the galaxy
merger rate in various ways, and find that a linear form
of $R_{\rm g} = C\times z + R_{\rm g,0}$ gives the best fit
(as shown by the solid line in Figure~13).    However, as can be seen in
Figure~13, this linear fit predicts that the merger rate should drop
quicker at $z < 0.2$ than what is actually seen. This is potentially
the result of either the merger rate dropping linearly from $z \sim 3$
until $z \sim 0.7$, and then dropping very quickly and remaining
relatively constant at $z < 0.6$.  It also could be the result of
the time-scale for the pair measurements at $z \sim 0$ to be incorrect.
If the time-scale were 0.4 Gyr, this would drop the merger rate
by a factor of two for the $z \sim 0$ point, although this would still
not alleviate the problem of matching with the rate of linear decrease
seen at higher redshifts. The merger rates at $z < 0.5$ appear to
be constant for M$_{*} > 10^{10}$ \solm galaxies.

The merger rate therefore decreases nearly linearly with redshift at 
$0.7 < z < 3$, whereas
the merger fraction is relatively well fit as a power-law decrease over
the same epoch.  While the galaxy number densities for M$_{*} > 10^{10}$ \solm
systems remains relatively constant at $z < 1$, they decline quickly
at higher redshifts (e.g., Conselice 2007).  This is matched, and then
some, by the increased merger fraction, which produces a linear decline
with decreasing redshift.  It is possible, as for the case of
very massive galaxies discussed in Bluck et al. (2009) for the
number density decline to match the increased merger fraction, producing
a relatively flat evolution of the merger rate.

The time integral of the merger rate gives us the number of mergers which
have occurred per unit volume since $z = 3$.  The result of this calculation
for the number of mergers which have occurred in a 10 Mpc$^{3}$ co-moving
cube is shown in Figure~14.  The solid lines show the evolution for the
same time-scales as in Figure~12.  Clearly, the number of mergers in
this volume, which contains roughly six galaxies today, increases rapidly at
higher redshifts, but levels off at $z < 1$.  Most of the merging in the
universe, as found in previous work (e.g., Conselice et al. 2003;
Lotz et al. 2008a), is at higher redshifts.  As discussed in Bertone
\& Conselice (2009, submitted), this is in disagreement with Cold
Dark Matter semi-analytical models, such that we are finding a higher
merger rate at very high redshifts, compared with the simulations.

\begin{figure}
 \vbox to 140mm{
\includegraphics[angle=0, width=90mm]{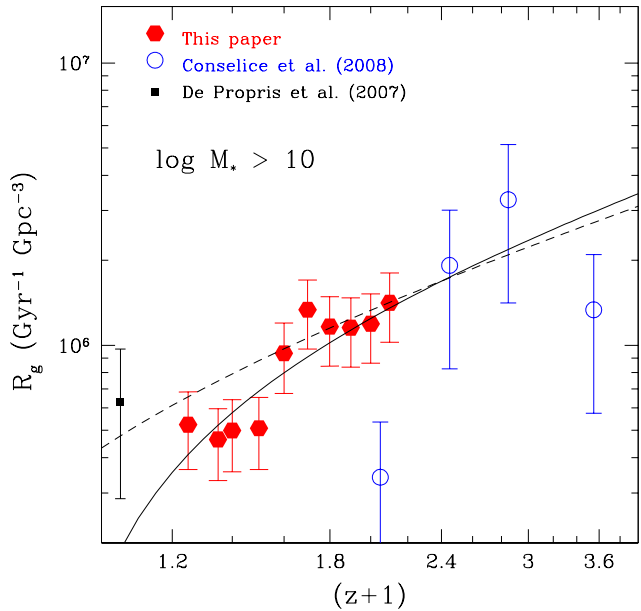}
 \caption{The galaxy merger rate ($\Re_{\rm g}$), defined as
the number of galaxies undergoing a merger per unit time, per
unit co-moving volume, as a function of redshift.  Shown here
are only systems with M$_{*} > 10^{10}$ \solm, as measured
by the CAS system from this paper at $0.2 < z < 1.2$; 
Conselice et al. (2008a) for $z > 0.6$ galaxies (blue circles),
while the $z = 0.05$ is the rate derived from De Propris et al.
(2007).}
} \label{sample-figure}
\end{figure}

\begin{figure}
 \vbox to 140mm{
\includegraphics[angle=0, width=90mm]{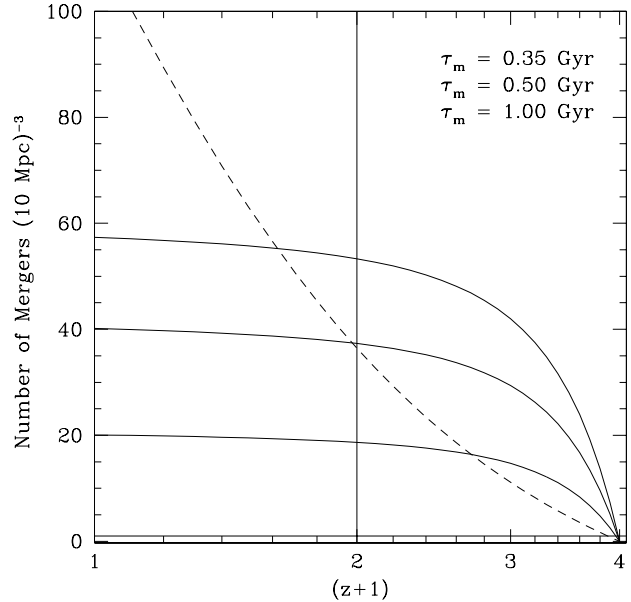}
\caption{The integral of the galaxy merger rate ($\Re_{\rm g}$),
from $z = 3$ to $z = 0$ over a volume of (10 Mpc)$^{3}$.  
This integral gives the total integrated number of mergers which
have occurred within this volume since $z = 3$.  As for the integral
of $\Gamma^{-1}$ (Figure~12), most of this merging occurs within a co-moving
volume at higher redshifts, with roughly 20-60 mergers occurring at $z < 3$.}
} \label{sample-figure}
\end{figure}

\section{Summary}

We have carried out an analysis of the structural CAS parameters
for a sample of $>$ 20,000 galaxies with stellar masses M$_{*} > 10^{10}$
\solm within the EGS and COSMOS fields, between $z = 0.2$ and
$z = 1.2$.  We explore in this paper the merger fraction history,
including various parameterisations, comparison of structural and
pair merger fractions, the merger rate and role of mergers in galaxy
formation, as well as systematics which are possibly playing a
role in the derived merger fraction.  The primary results from this paper
are:

\noindent I.  We find through the CAS structural method that the
merger fraction slightly increases from $z = 0.2$ to $z = 1.2$ with
the merger fraction increasing from $f_{m} = 0.04\pm0.01$ to 
$f_{m} = 0.13\pm0.01$.  

\noindent II. We compare our derived merger fractions to previously
published merger fractions from Lotz et al. (2008a) and pair studies
from Kartaltepe et al. (2007), Lin et al. (2008) and de Ravel et al. (2008).
Our results are comparable with the Gini/M$_{20}$ derived merger
fractions from Lotz et al. (2008a), but we find our merger fractions
are between 3-6 times higher than those derived from pairs, even
within the same stellar mass selection.  We argue
this in some detail through the use of a M$_{*} > 10^{11}$ \solm
sample of galaxies in the EGS.  We also show that this ratio of
structural mergers and pair fractions is however predicted in the latest
N-body models of galaxy mergers from Lotz et al. (2008b) and is
due to differing merger time-scales.

\noindent III. We investigate various methods, including the use of priors,
for parameterising the merger history through the CAS method.  We 
show that the power-law formalism, whereby the merger fraction is
parameterised by $f (z) = f_{0} \times (1+z)^{m}$, is a poor
fit to the merger fractions at $z < 1$.  We further show that
the value of the power-law slope, $m$, can vary depending on whether
a prior is used to set the local $z = 0$ merger fraction.  If we fit
only our $z = 0.2-1.2$ merger fractions from COSMOS+EGS we fit
a merger fraction evolution of $f(z) = 0.025\pm0.005(1+z)^{2.3\pm0.4}$. 
However,
by using a prior of $f_0 = 0.009$ from de Propris et al. (2007),
the slope of the power-law fit is $m = 3.8\pm0.2$

\noindent IV. We find that a combined power-law exponential is 
a better fit to the merger fraction at $z < 3$ than just a simple 
power-law.  We fit this form to the merger fraction using a $z = 0$ 
prior, and without, finding a similar form that predicts a turnover in the
merger fraction for M$_* > 10^{10}$ \solm galaxies at z = 2.

\noindent V. We calculate the merger rate by using the latest
measures of merger time-scales from merger simulations, and number
densities for galaxies with M$_* > 10^{10}$ \solm from
Conselice et al. (2007b).  Although the errors on measuring
the merger rate are large, resulting from uncertainties 
in the merger fraction, the merger time-scale,
and the galaxy density, it is the ultimate quantity, and can
reveal the role of mergers in forming galaxies.  We find
that the merger rate is roughly constant up to $z = 0.7$, and
increases at $z > 1$ by a factor of 10. We also show that the merger
rate for  M$_* > 10^{10}$ \solm galaxies can be parameterised up to
$z \sim 2$ as a linear increasing function.

We thank the staff at the Palomar and Keck observatories for invaluable
assistance in collecting the data used for the Extended Groth Strip survey.  
Funding to support this effort came from a National Science Foundation
Astronomy \& Astrophysics Fellowship, and grants from the Science Technology
Facilities Council (STFC).
Support for the ACS imaging of the EGS in GO program 10134 was 
provided by NASA through NASA grant HST-G0-10134.13-A from the Space 
Telescope Science Institute, which is operated by the Association of 
Universities for Research in Astronomy, Inc., under NASA contract NAS 5-26555.

\appendix

\label{lastpage}

\end{document}